\documentclass[twocolumn]{aa}
\usepackage{epsfig}
\usepackage{graphicx}
\usepackage{lscape}

\def\lsim{\,\lower2truept\hbox{${<\atop\hbox{\raise4truept\hbox{$\sim$}}}$}\,}
\def\gsim{\,\lower2truept\hbox{${> \atop\hbox{\raise4truept\hbox{$\sim$}}}$}\,}

\begin{document}
   \title{Constraints on Evolutionary Properties of GHz Peaked Spectrum Galaxies}

      \author{S. Tinti
          \inst{1} 
          \and G. De Zotti \inst{2,1}
          }

   \offprints{S. Tinti,\email{tinti@sissa.it}}

\institute{SISSA/ISAS, Via Beirut 4, 34014, Trieste, Italy  
\and INAF, Osservatorio Astronomico, Vicolo
dell'Osservatorio 5, 35122, Padova, Italy  
}

   \date{...}

   \abstract{We have used the available samples of Gigahertz Peaked
Spectrum (GPS) galaxies to investigate their evolution properties
in the framework of the ``youth'' scenario. Care was taken to
properly allow for the different selection criteria used to define
the samples. We find that the observed redshift and peak frequency
distributions can be satisfactorily accounted for in terms of
simple luminosity evolution of individual sources, along the lines
discussed by Fanti et al. (1995) and Begelman (1996, 1999),
although the derived parameter values have large uncertainties due
to ambiguities in the selection of GPS sources and to the
incompleteness of redshift measurements. However the simplest
self-similar model, whereby the evolution is controlled only by
the radial profile of the density of the ambient medium is not
good enough and one additional parameter needs to be introduced.
The fit requires a decrease of the emitted power and of the peak
luminosity with source age or with decreasing peak frequency, at
variance with the Snellen et al. (2000) model. On the other hand,
our analysis confirms the rather flat slope of the luminosity
function, found by Snellen et al. (2000) who also report
indications of a high luminosity break, not required by the data
sets we have used. Our results suggest that the GPS galaxies are
the precursors of extended radio sources with luminosities below
the break of the luminosity function. No cosmological evolution of
the GPS galaxy population is required by presently available data.

   }

   \maketitle
%

\section{Introduction}

Several lines of evidence increasingly point to Gigahertz Peaked
Spectrum (GPS) sources (see O'Dea 1998 for a comprehensive review)
being extremely young. Spectral age constraints, energy supply
arguments and, most convincingly, VLBI kinematic studies, are all
consistent with the current radio activity in these objects having
turned on less than about 1000 years ago (Conway 2002), although
in some cases there is evidence that this is not the first
instance of activity, but sources appear to have been re-activated
after a period of quiescence (Baum et al. 1990), perhaps by effect
of merger activity (Tingay et al. 2003). These sources  thus offer
the exciting opportunity of studying radio sources in early stages
of development.

There is a clear anti-correlation between the peak (turnover)
frequency, $\nu_p$, and the projected linear size of GPS sources
(Fanti et al. 1990; O'Dea \& Baum 1997). Although this
anti-correlation does not necessarily define the evolutionary
track, a decrease of the peak frequency as the emitting blob
expands is indicated, suggesting a scenario whereby GPS sources
evolve into Compact Steep Spectrum (CSS) sources and, finally,
into the classical large scale radio sources (Fanti et al. 1995;
Readhead et al. 1996; Begelman 1996; Snellen et al. 2000).

The self-similar evolution model by Begelman (1996, 1999) imply
that the evolution of the radio power as the source expands,
depends on the radial profile of the external density. To test
this scenario and to constrain the birth rates and the
evolutionary properties of these sources, detailed comparisons of
the model predictions with survey data are necessary. An
exploratory study in this direction was presented by De Zotti et
al. (2000), who showed that this kind of models may account for
the observed counts, redshift and peak frequency distributions of
the samples then available, but for rather unexpected values of
the parameters.

Since then, new important surveys of GPS sources have been
published, extending the coverage of the peak frequency-redshift
plane. The Dallacasa et al. (2000) and Bolton et al. (2004)
samples comprise sources with high turnover frequencies (called
High Frequency Peakers, HFPs), which are presumably the youngest.
Snellen et al. (2002) and Edwards \& Tingay (2004) selected
samples of southern GPS sources, thus improving the sky coverage.
Snellen et al. (2004) focussed on the relatively rare low-$z$
sources. The first year WMAP survey (Bennett et al. 2003b) has
provided a complete sample of over 200 sources with simultaneous
flux density measurements at 4 frequencies (22.8, 33.0, 40.7,
60.8, and 93.5 GHz), that can be exploited to search for extreme
GPS sources, peaking at mm wavelengths.

Moreover, the variability study of the Dallacasa et al. (2000)
sample, carried out by Tinti et al. (2005), has shown that most
HFP candidates classified as quasars are most likely blazars
caught by during a flare of a highly self-absorbed component
dominating the emission, while candidates classified as galaxies
are consistent with being bona fide HFPs. Polarization
measurements (Dallacasa et al., in preparation) lend further
support to this conclusion. Torniainen et al. (2005) from a study
of the long term variability of 33 objects previously classified
as GPS or HFP sources, mostly identified with quasars and mostly
peaking above 5 GHz, concluded that only 5 keep the GPS properties
over time. Clearly, a serious blazar contamination of GPS samples
may lead astray analyses of their statistical properties.

In view of these new data, the previous conclusions on the
evolution of the GPS population need to be reconsidered in depth.
In this paper we present a new analysis, still in the framework of
Begelman's (1996, 1999). Since, as mentioned above, there are
strong evidences that samples of GPS sources identified as quasars
are heavily contaminated by flaring blazars, we confine our
analysis to GPS galaxies. Our reference evolutionary scenario is
briefly described in Sect.~2, in Sect.~3 we give an account of the
data sets we use, in Sect.~4 we analyze the effect of the
different selection criteria adopted to define the samples, in
Sect.~5 we present our results, in Sect.~6 we summarize and
discuss our main conclusions.

We have adopted a flat $\Lambda$CDM cosmology with
$\Omega_{\Lambda}=0.73$ and
$H_0=70\,\hbox{km}\,\hbox{s}^{-1}\,\hbox{Mpc}^{-1}$ (Bennett et
al. 2003a).

\section{The evolutionary scenario \label{sect:scenario}}

Following De Zotti et al. (2000), we make the following
assumptions:

\begin{enumerate}

\item The initial radio luminosity function (in units of Mpc$^{-3}\,
{\mathrm{d}\log L_i}^{-1}$) is described by a power law:
\begin{equation}
n(L_i) \propto \left({L_i\over L_\star}\right)^{-\beta}, \quad
L_{i,{\rm min}} \leq L_{i} \leq L_{i,{\rm max}},
\end{equation}
where $L_i$ is the luminosity before self-absorption.

\item In the GPS phase, the properties of the sources are determined
by the interaction of a compact, jet-driven, overpressured, non
thermal radio lobe with a dense interstellar medium (Begelman
1996, 1999; Bicknell et al. 1997); the timescale of the
interaction is very short in comparison with the
cosmological-expansion timescale, so that the luminosity evolution
of individual sources occurs at constant $z$. As the radio lobe
expands in the surrounding medium, the emitted radio power varies
with the source age, $\tau$, as $L_i \propto \tau^{-\eta}$, and
its linear size $l$ varies as $l \propto \tau^{\epsilon}$. If the
density of the surrounding medium scales with radius as $\rho_e
\propto r^{-n}$, we have (Begelman 1996, 1999) $\eta =
(n+4)/[4(5-n)]$ and $\epsilon = 3/(5-n)$. The Begelman's dependence of 
the radio power on source age assumes that the synchrotron emission is
dominated by the high pressure region close to the head of the jet.
Alternatively, Snellen et al.~(2000) and Alexander (2000) assume that the
emission comes from the full volume of the expanding cocoon, so that 
$\eta=(8-7n)/4(5-n)$.  
In Begelman's model the luminosity decreases with increasing source size
for $n\geq0$.  On the other hand, according to Snellen et
al. (2000) and Alexander (2000) the radio luminosity
increases as $l^{2/3}$ if $n=0$ (i.e. within the galaxy core radius) 
and then decreases as $l^{-0.5}$ as the density profile steepens to $n=2$ 
at larger radii.
Evolutionary tracks where the luminosity increases within
the core radius and then decreases in a falling atmosphere in the
power-diameter plane are also presented by Carvalho \& O'Dea
(2003) based on the results of a series of hydrodynamical simulations.
There is a clear
anticorrelation between intrinsic turnover frequency, $\nu_p$, and
linear size (O'Dea \& Baum 1997): $\nu_p \propto l^{-\delta}$,
with $\delta \simeq 0.65$. It follows that $\nu_p$ scales with
time as $\nu_p \propto \tau^{-\lambda}$, with $\lambda =
\delta\epsilon$.

\item The spectra of GPS sources are described by:
\begin{equation}
L_\nu = L_p\times \left\{ \begin{array}{ll}
       (\nu/\nu_p)^{\alpha_a} & \mbox{if $\nu < \nu_p$} \\
       (\nu/\nu_p)^{-\alpha}  & \mbox{if $\nu > \nu_p$}
       \end{array}
       \right.
\end{equation}
with $\alpha_a=0.8$ and $\alpha=0.75$, the mean values found by
Snellen et al. (1998b).

\end{enumerate}

\noindent As the radio lobe expands, the peak luminosity $L_p$
varies as the consequence of the variation of both the emitted
radio power and of $\nu_p$. Hence:
\begin{equation}
L_p(\nu_p) = L_{p,i} \tau^p = L_{p,i} \left({\nu_p \over
\nu_{p,i}}\right)^ {-p/\lambda} = L_{p,i} \left({\nu_p \over
\nu_{p,i}}\right)^ {\eta/\lambda - \alpha}  \label{eq:Lp}
\end{equation}
with $L_{p,i}(z) \propto L_{i}(z)$ and $p=-\eta+\alpha\lambda$.

If the birth rate of GPS sources is constant on time scales much
shorter than the cosmological-expansion timescale, the peak
luminosity function per unit $\mathrm{d}\log L_p$ is:
\begin{equation}
n(L_p) \propto L_p^{1/p}.
\end{equation}
Also, since, in this case, the {\it comoving} number of sources of
age $\tau$ within $\mathrm{d}\tau$ is simply proportional to
$\mathrm{d}\tau$ and $\mathrm{d}\tau/\mathrm{d}\nu_p \propto
(\nu_p/\nu_{p,i})^{-(1+1/\lambda)}$, the epoch dependent
luminosity function at a given frequency $\nu$
(Mpc$^{-3}\,{\mathrm{d}\log L_\nu}^{-1}\,\mathrm{GHz}^{-1}$)
writes:
\begin{equation}
n(L_\nu, \nu_p,z) = n_0 \left({L_{p,i}(L_\nu,\nu_p)\over
 L_\star(z)}\right)^{-\beta} \left({\nu_p\over \nu_{p,i}}\right)^{-(1+
1/\lambda)}\ , \label{eq:FL}
\end{equation}
where $L_\star(z)$ is the redshift-dependent normalization
luminosity. We have assumed luminosity evolution and adopted a
very simple parameterization for it:
\begin{equation}
L_\star(z) = L_0\times \left\{ \begin{array}{ll}
       (1+z)^k & \mbox{if $z < z_c$} \\
       (1+z_c)^k  & \mbox{if $z > z_c $}
       \end{array}
       \right. \label{eq:evol}
\end{equation}
The redshift $z_c$ at which luminosity evolution levels off is a
model parameter. We have normalized monochromatic luminosities to
$L_0=10^{32}\,\hbox{erg}\, \hbox{s}^{-1}\,\hbox{Hz}^{-1}$.

The luminosity function at a frequency $\nu_2$ is related to that
at a frequency $\nu_1$ by:
\begin{equation}
n(L_{\nu_2}) = n(L_{\nu_1})\times \left\{ \begin{array}{ll}
       (\nu_2/\nu_1)^{-\alpha_a} & \mbox{if $\nu_1< \nu_2 < \nu_p$} \\
       (\nu_1/\nu_p)^{\alpha_a} (\nu_2/\nu_p)^{\alpha} & \mbox{if
$\nu_1 < \nu_p < \nu_2$} \\
       (\nu_2/\nu_1)^{\alpha} & \mbox{if $\nu_p< \nu_1 < \nu_2$}
       \end{array}
       \right.
\end{equation}
The number counts per steradian of GPS sources brighter than
$S_\nu$ at the frequency $\nu$, with an observed peak frequency
$\max(\nu,\nu_{p,\rm{min}}) < \nu_{p,0} < \nu_{p,\rm{max}}$ are
given by
\begin{eqnarray}
&N&(>S_\nu; \nu_{p,0} >\nu) =
\int_0^{\min\left[z_f,z_m(S_\nu)\right]} \mathrm{d}z\,
{\mathrm{d}V\over \mathrm{d}z} \times \nonumber \\
&\times&
\int_{\max[\nu_{p,\rm{min}}(1+z),\nu(1+z)]}^{\min[\nu_{p,\rm{max}}
 (1+z),\nu_{p,i}]} \mathrm{d}\nu_p    \times \nonumber \\
&\times&  \int_{\log L_{\rm min}(S_\nu,z,\nu_p)}^{\log L_{\rm
max}(z)} \mathrm{d}\log L_\nu\, n(L_\nu,\nu_p,z) \, ,
\end{eqnarray}
where $z_f$ is the redshift of formation of the first GPS sources,
$z_m$ is the maximum redshift at which sources can have a flux
$\geq S_\nu$, $L_{\rm min}$ is the minimum luminosity of a source
of given $z$ and $\nu_p$ yielding a flux $\geq S_\nu$,
$\mathrm{d}V/\mathrm{d}z$ is the {\it comoving} volume element
within a solid angle $\omega$:
\begin{equation}
{\mathrm{d}V \over \mathrm{d}z} = {c\over H_0}\omega {d_L^2\over
(1+z)^2E(z)}
\end{equation}
$d_L(z)$ being the luminosity distance:
\begin{equation}
d_L(z) = {c\over H_0} (1+z)\int_0^z\frac{dz'}{E(z')}
\end{equation}
\begin{equation}
E(z)\equiv \left(\Omega_M(1+z)^3+\Omega_\Lambda\right)^{1/2} .
\end{equation}
The flux density at the frequency $\nu$ is related to the
rest-frame luminosity $L_\nu$ by:
\begin{equation}
S_\nu = {L_\nu K(z) \over 4\pi d_L^2}
\end{equation}
\begin{equation}
K(z)=(1+z) {L_{\nu(1+z)}\over L_\nu} \, .
\end{equation}
$K(z)$ being the K-correction.

Similarly, the number counts of GPS sources with an observed peak
frequency $\nu_{p,\rm{min}} < \nu_{p,0} <
\min(\nu,\nu_{p,\rm{max}})$ are given by
\begin{eqnarray}
&N&(>S_\nu; \nu_{p,0} < \nu) =
\int_0^{\min\left[z_f,z_m(S_\nu)\right]} \mathrm{d}z\,
{\mathrm{d}V\over \mathrm{d}z} \times \nonumber \\
&\times&
\int_{\nu_{p,\rm{min}}(1+z)}^{\min[\nu\,(1+z),\nu_{p,\rm{max}}(1+z),\nu_{p,i}]}
\mathrm{d}\nu_p    \times \nonumber \\
&\times&  \int_{\log L_{\rm min}(S_\nu,z,\nu_p)}^{\log L_{\rm
max}(z)} \mathrm{d}\log L_\nu\, n(L_\nu,\nu_p,z) \, ,
\end{eqnarray}
The distribution of observed peak frequencies per unit
$\mathrm{d}\nu_{p,0}$ in a flux limited sample, ${\cal
N}(\nu_{p,0};>S_\nu)$, is given by:
\begin{eqnarray}
&{\cal N}&(\nu_{p,0};>S_\nu) =
 \int_0^{\min\left[z_f,z_m(S_\nu,\nu_{p,0})\right]} \mathrm{d}z\,
{\mathrm{d}V\over \mathrm{d}z} \times \nonumber \\
&\times& \int_{\log L_{\rm min}(S_\nu,z,\nu_{p,0}(1+z))} ^{\log
L_{\rm max}(z,\nu_{p,0}(1+z))}\!\!\!\!\!\!\!\! \mathrm{d}\log
L_\nu\, n[L_\nu,z,\nu_{p,0}(1+z)].
\end{eqnarray}

\begin{table}
\begin{center}
\caption{Summary of GPS samples.} \label{samples}
\begin{tabular}{lllllll}
\hline\hline Sample & $S_{\rm lim}$ & $\nu_o$& area   &
$\nu_{p}^{\rm min}$& $\nu_{p}^{\rm max}$&
$N_{\rm tot}^{\rm gal}$\\
       &  Jy           & GHz    & deg$^2$ & GHz            & GHz            & \\
\hline
HFP    & 0.3           &  4.9   & 15840  &  5.34 &    11.1        & 5  \\
Stan   & 1.0           &  5.0   & 24600  &  0.4  &    4.9         & 19 \\
B\_SA  & 0.025         &  15.0  & 176    &  3.43 &    127.0       & 10 \\
B\_SB  & 0.060         &  15.0  & 70     &  3.29 &    10.3        & 0  \\
Snel   & 0.018         &  0.325 & 522    &  1.0  &    5.7         & 14 \\
CORA   & 0.1           &  1.4   & 2850   &  0.460&    2.3         & 6  \\
Park   & 0.5           &  2.7   & 12802  &  0.4  &    5.0         & 48  \\
Edwa   & 0.95          &  5.0   & 16414  &  0.7  &    7.5         & 9  \\
\hline
\end{tabular}
\end{center}
\end{table}

\section{GPS samples \label{sect:samples}}

There is no clear-cut definition of GPS sources and different
criteria have been adopted for identifying them. On the other
hand, we need to combine at least the most reliable samples to
gather enough data for a meaningful analysis to be possible. We
now describe the samples used and the corrections applied to make
them as homogeneous as possible. The main properties of the
samples are summarized in Table~\ref{samples}, where HFP, Stan,
B\_SA, B\_SB, Snel, CORA, Park, and Edwa denote, respectively,
the samples by Dallacasa et al. (2000, HFP sample), by
Stanghellini et al. (1998), by Bolton et al. (2004, 9C samples A
and B), by Snellen et al. (1998, as revised by Snellen et al.
2000, faint WENSS sample), by Snellen et al. (2004, CORALZ
sample), by Snellen et al. (2002, Parkes sample), and by Edwards
\& Tingay (2004, ATCA sample).

\begin{table}[t]
\begin{center}
\caption{GPS candidates from Bolton et al. 2004.} \label{bolton}
\begin{tabular}{lclll}
 & & & &  \\ \hline \hline
Name       &    &  & $S_p$       &$\nu_p$ \\
           &        &  & (mJy)       & (GHz)  \\ \hline
J$0003+2740^A$ &  Q? &        &  $76  \pm 1  $  &  $5.4  \pm 0.1 $ \\
J$0003+3010^A$ &  G? & 0.55   &  $60  \pm 3  $  &  $10.3 \pm 0.3 $ \\
J$0010+2854^A$ &  Q? &        &  $100 \pm 5  $  &  $52   \pm 6   $ \\
J$0012+3353^A$ &  G? & 0.50   &  $199 \pm 12 $  &  $127  \pm 36  $ \\
J$0012+3053^A$ &  G? & 0.45   &  $27  \pm 1  $  &  $20   \pm 4   $ \\
J$0020+3152^A$ &  -  & 1.1    &  $43.8\pm 0.8$  &  $4.91 \pm 0.09$ \\
J$0024+2911^A$ &  Q? &        &  $42  \pm 2  $  &  $13.9 \pm 0.5 $ \\
J$0032+2758^A$ &  G  & 0.51   &  $34.4\pm 0.4$  &  $4.6  \pm 0.2 $ \\
J$0919+3324^B$ &  G? & 0.35   &  $430 \pm 16 $  &  $10.3 \pm 0.5 $ \\
J$0925+3127^B$ &  G? & 0.26   &  $135 \pm 2  $  &  $3.29 \pm 0.06$ \\
J$0931+2750^B$ &  G? & 0.49   &  $169 \pm 3  $  &  $8.3  \pm 0.3 $ \\
J$0935+2917^A$ &  Q? &        &  $47.3\pm 0.9$  &  $5.7  \pm 0.2 $ \\
J$0936+3207^A$ &  G  & 0.20   &  $56  \pm 5  $  &  $14   \pm 1   $ \\
J$0936+2624^B$ &  Q? &        &  $179 \pm 18 $  &  $3.8  \pm 0.3 $ \\
J$0940+2603^B$ &  G? & 0.37   &  $495 \pm 10 $  &  $6.1  \pm 0.8 $ \\
J$0945+3534^B$ &  Q? &        &  $413 \pm 8  $  &  $4.0  \pm 0.2 $ \\
J$0952+3512^B$ &  Q? &        &  $427 \pm 8  $  &  $4.1  \pm 0.2 $ \\
J$0955+3335^B$ &  Q? &        &  $107 \pm 2  $  &  $5.9  \pm 0.2 $ \\
J$1506+4239^A$ &  G  & 0.38   &  $766 \pm 19 $  &  $14   \pm 1   $ \\
J$1517+3936^A$ &  G  & 0.46   &  $43  \pm 2  $  &  $21   \pm 3   $ \\
J$1521+4336^A$ &  Q? &        &  $431 \pm 10 $  &  $6.2  \pm 0.2 $ \\
J$1526+3712^A$ &  G? & 0.61   &  $74  \pm 2  $  &  $6.9  \pm 0.3 $ \\
J$1526+4201^A$ &  G  & 0.36   &  $67  \pm 1  $  &  $8.1  \pm 0.2 $ \\
J$1528+3816^A$ &  G? & 0.52   &  $80  \pm 3  $  &  $26   \pm 5   $ \\
J$1530+3758^A$ &  G  & 0.19   &  $141 \pm 3  $  &  $3.43 \pm 0.06$ \\
J$1540+4138^A$ &  G  & 0.17   &  $46  \pm 8  $  &  $8.8  \pm 0.2 $ \\
J$1550+4536^A$ &  G  & 0.50   &  $62  \pm 1  $  &  $3.5  \pm 0.07$ \\
J$1554+4350^A$ &  -  & 1.2    &  $45.0\pm 0.8$  &  $10.9 \pm 0.3 $ \\
J$1554+4348^A$ &  G? & 0.36   &  $63  \pm 1  $  &  $3.7  \pm 0.1 $ \\
J$1556+4259^A$ &  Q? &        &  $94  \pm 4  $  &  $4.3  \pm 0.2 $ \\
\hline
\end{tabular}
\end{center}
\end{table}

\subsection{The 9C samples}

Bolton et al. (2004) selected two complete samples of sources from
the first three regions of the 9C survey: a deeper sample (sample
A), complete to 25 mJy and containing 124 sources over a total
area of 176 square degrees; a shallower sample (sample B),
complete to 60 mJy, comprising 70 sources in an area of 246 square
degrees, including the area covered by sample A. Restricting the
sample B to the 70 sq. deg. not overlapping the sample A we are
left with 31 sources. Simultaneous observations of each source
were made at frequencies of 1.4, 4.8, 22 and 43 GHz with the VLA
and at 15 GHz with the Ryle Telescope. In addition, 51 sources
were observed within a few months at 31 GHz with the Owens Valley
Radio Observatory (OVRO) 40m telescope. Sources with spectral
index between 1.4 and 4.8 GHz $\alpha_{1.4}^{4.8}<-0.1$ ($S_\nu
\propto \nu^{-\alpha}$) were referred to as GPS sources. The
adopted criterion implies that sources peaking at $\nu_p \gsim
5\,$GHz were preferentially (but not exclusively) selected. With
this definition, there are 22 GPS sources in sample A (14
galaxies, 6 quasars and 2 unidentified sources) and 8 (4 galaxies
and 4 quasars) in the (redefined) sample B.

In order to estimate the peak flux densities, $S_p$, and
frequencies, $\nu_p$, of the sources, we have fitted the radio
spectra with the hyperbolic function used by Tinti et al. (2005):
\begin{equation}
\log(S)=\log(S_p)+
b-\left[b^2+c^2(\log\nu-\log\nu_{p})^2\right]^{1/2}.
\label{eqhyp1}
\end{equation}
The best fit values of $\nu_p$ and $S_p$, obtained minimizing the
chi-square function with the Minuit package (CERN libraries), are
reported with their errors in column 4 and 5 of
Table~\ref{bolton}. 

No redshift measurements are available for these sources. However,
as shown by Snellen et al. (1996, 2002), GPS galaxies show a well
defined $R$-band Hubble diagram, with a low dispersion. In terms
of Gunn $r$ magnitudes, Snellen et al. (1996) found a best fit
relation:
\begin{equation}
r=22.7+7.4\log(z). \label{hubble_rel}
\end{equation}
The conversion from the Kron-Cousins $R_c$ magnitudes measured by
Bolton et al. (2004) to the Gunn $r$ magnitudes was made taking
$r=R_c+0.3$ (Fukugita et al. 1995). The redshift estimates are
given in Table~\ref{bolton}. We do not count as galaxies the
optically point-like objects, classified as G? by Bolton et al.
(2004) because of their red colours ($O-R \ge 1.6$): if they were
bright galaxies at the redshifts estimated from
Eq.~(\ref{hubble_rel}) they should be resolved. This leaves 10 GPS
galaxies in sample A and zero in sample B (see
Table~\ref{samples}).

It is important to note that other criteria for selecting GPS
samples include tighter constraints on the low- and/or
high-frequency spectral indices. The effect of the different
selection criteria is discussed in Sect.~\ref{sect:selection}.

\begin{table}[t]
\begin{center}
\caption{The HFP galaxy sample. The * denotes an empty field,
which was attributed $z=1.5$.} \label{hfp_gal}
\begin{tabular}{llll}
\hline\hline
Name       & z & $\nu_p$ \\
           &   & GHz      \\
\hline
 0428+3259 & 0.3     & 7.3  \\
 0655+4100 & 0.02156 & 7.8  \\
 1407+2827 & 0.0769  & 5.34 \\
 1511+0518 & 0.084   & 11.1 \\
 1735+5049 & 1.5*    & 6.4  \\
\hline
\end{tabular}
\end{center}
\end{table}

\subsection{The HFP sample}
The bright HFP sample by Dallacasa et al. (2000) was selected by
cross-correlating the 87GB (Gregory et al. 1996) sources with
$S_{4.9{\rm GHz}} \geq 300\,$mJy with the NVSS catalogue (Condon
et al. 1998) at 1.4 GHz and picking out those with inverted
spectra ($\alpha < -0.5$, $S\propto\nu^{-\alpha}$). It was then
``cleaned'' by means of simultaneous multifrequency VLA
observations, leaving 55 sources whose single-epoch radio spectrum
peaks at frequencies ranging from a few GHz to about 22 GHz. The
sample of HFP candidates comprises 11 galaxies (including a type 1
Seyfert), 36 quasars, and 8 still unidentified sources (Dallacasa
et al. 2002), over an area of $15,840\,\hbox{deg}^2$.

Although there are, in this sample, some sources with $\nu_p <
4.9\,$GHz, the selection criterion biases the sample against such
values of $\nu_p$ in a way that we are unable to quantify.
Therefore we have chosen to confine ourselves to sources with
$\nu_p > 4.9\,$GHz. Moreover we have excluded from the sample the
objects with 4.9 GHz flux densities smaller than the completeness
limit of 300 mJy when were they re-observed by Dallacasa et al.
(2000) and Tinti et al. (2005). After having applied these
additional constraints, we are left with 5 HFP galaxies, listed in
Table~\ref{hfp_gal}.

\begin{table}
\begin{center}
\caption{The faint GPS galaxies from WENSS. The * denotes empty
fields, which were attributed $z=1.5$. Values of $z$ with 3
significant digits are spectroscopic, the others are photometric
estimates.} \label{sne_gal}
\begin{tabular}{llll}
\hline\hline
Name       & z & $\nu_p$ \\
           &   & GHz      \\
\hline
B0400+6042 & 1.5*   &   1.0  \\
B0436+6152 & 1.5*   &   1.0  \\
B0535+6743 & 1.5    &   5.7  \\
B0539+6200 & 1.4    &   1.9  \\
B0830+5813 & 0.093  &   1.6  \\
B1525+6801 & 1.1    &   1.8  \\
B1551+6822 & 1.3    &   1.5  \\
B1557+6220 & 0.9    &   2.3  \\
B1600+7131 & 1.5*   &   1.7  \\
B1622+6630 & 0.201  &   4.0  \\
B1655+6446 & 1.5*   &   1.0  \\
B1841+6715 & 0.486  &   2.1  \\
B1942+7214 & 1.1    &   1.4  \\
B1946+7048 & 0.101  &   1.8  \\
\hline
\end{tabular}
\end{center}
\end{table}

\subsection{The faint GPS sample from WENSS}

The selection of this sample is described in detail in Snellen et
al. (1998). Snellen et al. (2000) applied stricter criteria
allowing a better control of selection effects; they kept only the
14 objects with inverted spectra between 325 and 5 GHz, and with
325 MHz flux densities  $> 20\,$mJy, over an area of
$522\,\hbox{deg}^2$. The redshift of identified sources, all
classified as galaxies, had to be estimated from their optical
magnitudes, using Eq.~(\ref{hubble_rel}). For the 4 unidentified
sources, also assumed to be galaxies, a redshift of z=1.5 was
assumed. The relevant data for all the 14 objects are given in
Table~\ref{sne_gal}.

\begin{table}
\begin{center}
\caption{The bright GPS galaxies from Stanghellini et al. (1998).
The * denotes photometric redshift estimates. } \label{sta_gal}
\begin{tabular}{llll}
\hline\hline
Name       & $z$ & $\nu_p$ \\
           &   & GHz      \\
\hline
0019-000   & 0.305 &  0.8 \\
0108+388   & 0.669 &  3.9 \\
0316+161   & 1.2*  &  0.8 \\
0428+205   & 0.219 &  1.0 \\
0500+019   & 0.583 &  2.0 \\
0710+439   & 0.518 &  1.9 \\
0941-080   & 0.228 &  0.5 \\
1031+567   & 0.459 &  1.3 \\
1117+146   & 0.362 &  0.5 \\
1323+321   & 0.369 &  0.5 \\
1345+125   & 0.122 &  0.6 \\
1358+624   & 0.431 &  0.5 \\
1404+286   & 0.077 &  4.9 \\
1600+335   & 1.1*  &  2.6 \\
1607+268   & 0.473 &  1.0 \\
2008-068   & 0.7*  &  1.3 \\
2128+048   & 0.99  &  0.8 \\
2210+016   & 1.0*  &  0.4 \\
2352+495   & 0.237 &  0.7 \\
\hline
\end{tabular}
\end{center}
\end{table}

\subsection{The bright GPS sample}

Stanghellini et al. (1998) selected candidate radio bright GPS
sources from the K\"uhr et al. (1981) catalogue ($S_{5{\rm
GHz}}\ge 1\,$Jy), over an area of about $24,600\,\hbox{deg}^2$.
The sample was then cleaned by means of multifrequency VLA and
WSRT observations, supplemented with literature data. They picked
out GPS candidates with a turnover frequency between 0.4 and 6
GHz, and an optically thin spectral index $\alpha_{\rm thin}>0.5$
($S_\nu \propto \nu^{-\alpha}$), beyond the peak. The final
complete sample consists of 33 GPS sources, 19 of which are
identified with galaxies. Four galaxies do not have spectroscopic
redshift; estimates by Snellen et al. (2000) from their optical
magnitudes [Eq.~(\ref{hubble_rel})] are denoted by a * in
Table~\ref{sta_gal}, where the relevant data for the 19 galaxies
are listed.

\begin{table}
\begin{center}
\caption{The ATCA sample of GPS galaxies from (Edwards \& Tingay
2004). The * denotes empty fields, which were attributed $z=1.5$.}
\label{atca_gal}
\begin{tabular}{lll}
\hline\hline
Name       & $z$ & $\nu_p$ \\
           &   & GHz  \\
\hline
J0241-0815  &   0.004 &  7.5 \\
J1543-0757  &   0.172 &  0.7 \\
J1658-0739  &   1.5*  &  4.8 \\
J1726-6427  &   1.5*  &  1.1 \\
J1723-6500  &   0.014 &  2.7 \\
J1744-5144  &   1.5*  &  1.0 \\
J1939-6342  &   0.183 &  1.4 \\
J2257-3627  &   0.006 &  2.7 \\
J2336-5236  &   1.5*  &  1.1 \\
\hline
\end{tabular}
\end{center}
\end{table}

\subsection{The ATCA GPS sample}

Edwards \& Tingay (2004) have used data from an Australia
Telescope Compact Array (ATCA) program of multi-frequency,
multi-epoch monitoring of the portion of the VSOP survey sample
(Hirabayashi et al. 2000) with declinations $< 10^\circ$. The
original sample is defined by: $S_{\rm 5GHz} > 0.95\,$Jy,
$\alpha<0.45$, $|b|> 10^\circ$. Taking into account the further
constraint $\delta < 10^\circ$, we estimate that the area covered
is $\sim 5\,$sr. The selected sources have $\alpha_{\rm thin}>0.5$
and spectral curvature $\alpha_{\rm thin}-\alpha_{\rm thick}>0.6$.

We have excluded from the sample the gravitationally lensed source
J$0414+0534$ because our models do not include the effect of
lensing, and J$1522-2730$, classified as a BL Lac object, whose
variability properties suggests that it is less likely to be a
truly GPS source (Edwards \& Tingay 2004). This leaves 7 GPS
galaxies (5 having spectroscopic redshifts), 16 quasars (15 with
spectroscopic redshift), and 2 empty fields (see
Table~\ref{atca_gal}).

\begin{table}
\begin{center}
\caption{The Parkes half-Jansky sample of GPS galaxies (Snellen et
al.~2002). The * denotes photometric redshift estimates. }
\label{par_gal}
\begin{tabular}{llll}
\hline\hline
Name       & $m_R$ & $z$ & $\nu_p$ \\
           &     &   & GHz      \\
\hline
J0022+0014 & 18.10 &    0.305    & 0.6\\
J0108-1201 & 22.39 &    1.0$^*$ & 1.0\\
J0206-3024 & 21.00 &    0.65$^*$ & 0.5\\
J0210+0419 & 24.1  &    1.5$^*$    & 1.3\\
J0210-2213 & 23.52 &    1.4$^*$ & 1.5\\
J0242-2132 & 17.10 &    0.314     & 1.0\\
J0323+0534 & 19.20 &    0.37$^*$ & 0.4\\
J0401-2921 & 21.0  &    0.65$^*$ & 1.0\\
J0407-3924 & 20.40 &    0.54$^*$ & 0.4\\
J0407-2757 & 21.14 &    0.68$^*$ & 1.5\\
J0433-0229 & 19.10 &    0.36$^*$ & 0.4\\
J0441-3340 & 21.0  &    0.65$^*$ & 1.2\\
J0457-0848 & 20.30 &    0.52$^*$ & 0.4\\
J0503+0203 & 21.0  &    0.583     & 2.5\\
J0943-0819 & 17.50 &    0.228     & 0.4\\
J0913+1454 & 20.0  &    0.47$^*$ & 1.1\\
J1044-2712 & 21.0  &    0.65$^*$ & 0.8\\
J1057+0012 & 21.0  &    0.65$^*$ & 1.6\\
J1109+1043 & 20.50 &    0.55$^*$ & 0.5\\
J1110-1858 & 19.60 &    0.497     & 1.0\\
J1120+1420 & 20.10 &    0.362     & 0.4\\
J1122-2742 & 21.0  &    0.65$^*$ & 0.8\\
J1135-0021 & 16.50 &    0.16$^*$ & 0.4\\
J1203+0414 & 18.80 &    0.33$^*$ & 0.4\\
J1345-3015 & 21.0  &    0.65$^*$ & 2.5\\
J1347+1217 & 15.20 &    0.122     & 0.4\\
J1350-2204 & 20.93 &    0.63$^*$ & 0.4\\
J1352+0232 & 20.00 &    0.47$^*$ & 0.4\\
J1352+1107 & 21.0  &    0.65$^*$ & 3.6\\
J1447-3409 & 21.00 &    0.65$^*$ & 0.5\\
J1506-0919 & 19.70 &    0.43$^*$ & 0.6\\
J1521+0430 & 22.10 &    1.296     & 1.0\\
J1543-0757 & 17.40 &    0.172     & 0.4\\
J1546+0026 & 20.10 &    0.556     & 0.6\\
J1548-1213 & 21.88 &    0.883     & 0.4\\
J1556-0622 & 22.20 &    0.94$^*$ & 0.4\\
J1604-2223 & 18.75 &    0.141     & 0.6\\
J1640+1220 & 21.36 &    1.150     & 0.4\\
J1648+0242 & 21.0  &    0.65$^*$ & 3.4\\
J1734+0926 & 20.80 &    0.61$^*$ & 1.0\\
J2011-0644 & 21.18 &    0.547     & 1.4\\
J2058+0540 & 23.40 &    1.381     & 0.4\\
J2123-0112 & 23.30 &    1.158     & 0.4\\
J2130+0502 & 22.21 &    0.990     & 1.0\\
J2151+0552 & 20.20 &    0.740     & 5.0\\
J2212+0152 & 22.0  &    0.88$^*$ & 0.4\\
J2325-0344 & 23.50 &    1.4$^*$ & 1.4\\
J2339-0604 & 22.91 &    1.2$^*$ & 0.4\\
\hline
\end{tabular}
\end{center}
\end{table}

\subsection{The Parkes half-Jy sample}

Snellen et al. (2002) selected from the Parkes multifrequency
survey data in a region of about 3.9 sr a southern/equatorial
sample of GPS sources with $S_{2.7{\rm GHz}}>0.5\,$Jy, excluding
objects identified as quasars. The sample (see
Table~\ref{par_gal}) consists of 49 objects with spectra peaking
at $\nu_p > 0.4\,$GHz, 38 of which are identified with galaxies,
10 are too faint to be identified and 1 is too close to a bright
star to allow identification, and is excluded from the statistical
analysis. The authors argue that, based on the magnitude
distribution of other GPS samples, the 10 faint sources are
unlikely to be quasars, and we assume them to be galaxies.
Spectroscopic redshifts are available for 18 objects. Estimates,
or lower limits, for the others have been obtained  through
Eq.~(\ref{hubble_rel}). No restrictions on either the low- or the
high-frequency spectral index are mentioned.

\begin{table}
\begin{center}
\caption{The CORALZ sample of local GPS galaxies (Snellen et
al.~2004).} \label{cora_gal}
\begin{tabular}{lll}
\hline\hline
Name       & $z$ & $\nu_p$ \\
           &   & GHz     \\
\hline
J073328+560541& 0.104&  0.460  \\
J073934+495438& 0.054&  0.950  \\
J083139+460800& 0.127&  2.200  \\
J090615+463618& 0.085&  0.680  \\
J131739+411545& 0.066&  2.300  \\
J171854+544148& 0.147&  0.480  \\
\hline
\end{tabular}
\end{center}
\end{table}

\subsection{The CORALZ sample}

The sample of Compact Radio sources at Low Redshift (CORALZ) was
selected by Snellen et al. (2004) solely on the basis of the radio
angular size ($\theta < 2''$), independent of radio spectra,
picking out sources associated with relatively bright galaxies. It
is estimated to be $\simeq 95\%$ complete for $S_{1.4{\rm GHz}}
>100\,$mJy over the redshift range $0.005 < z < 0.16$ in a region
of 2850 square degrees. The sample comprises 6 GPS galaxies with
$\nu_p > 0.4\,$GHz, all with spectroscopic redshifts
(Table~\ref{cora_gal}).

\begin{figure*}
\center{{\epsfig{figure=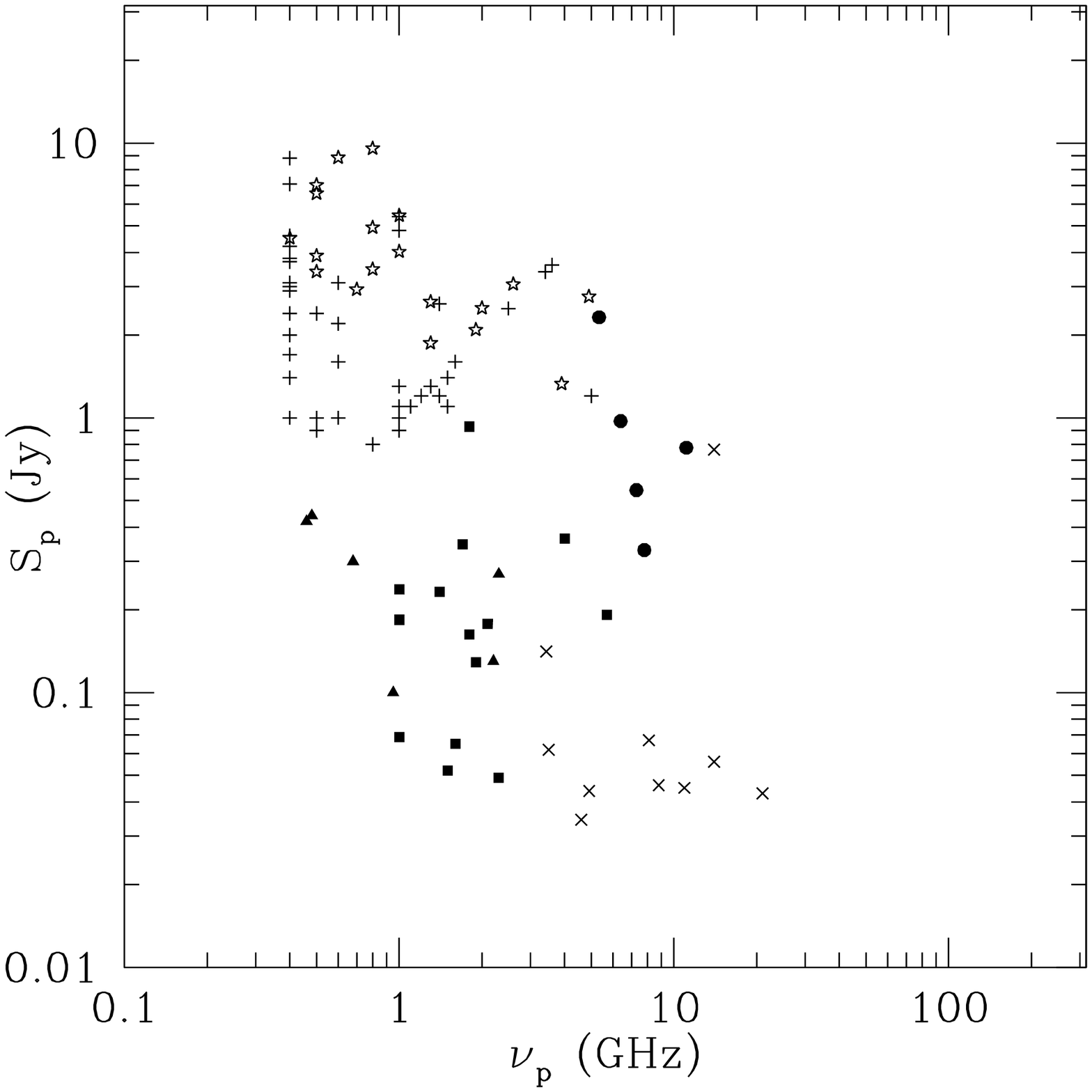,height=7cm}
\epsfig{figure=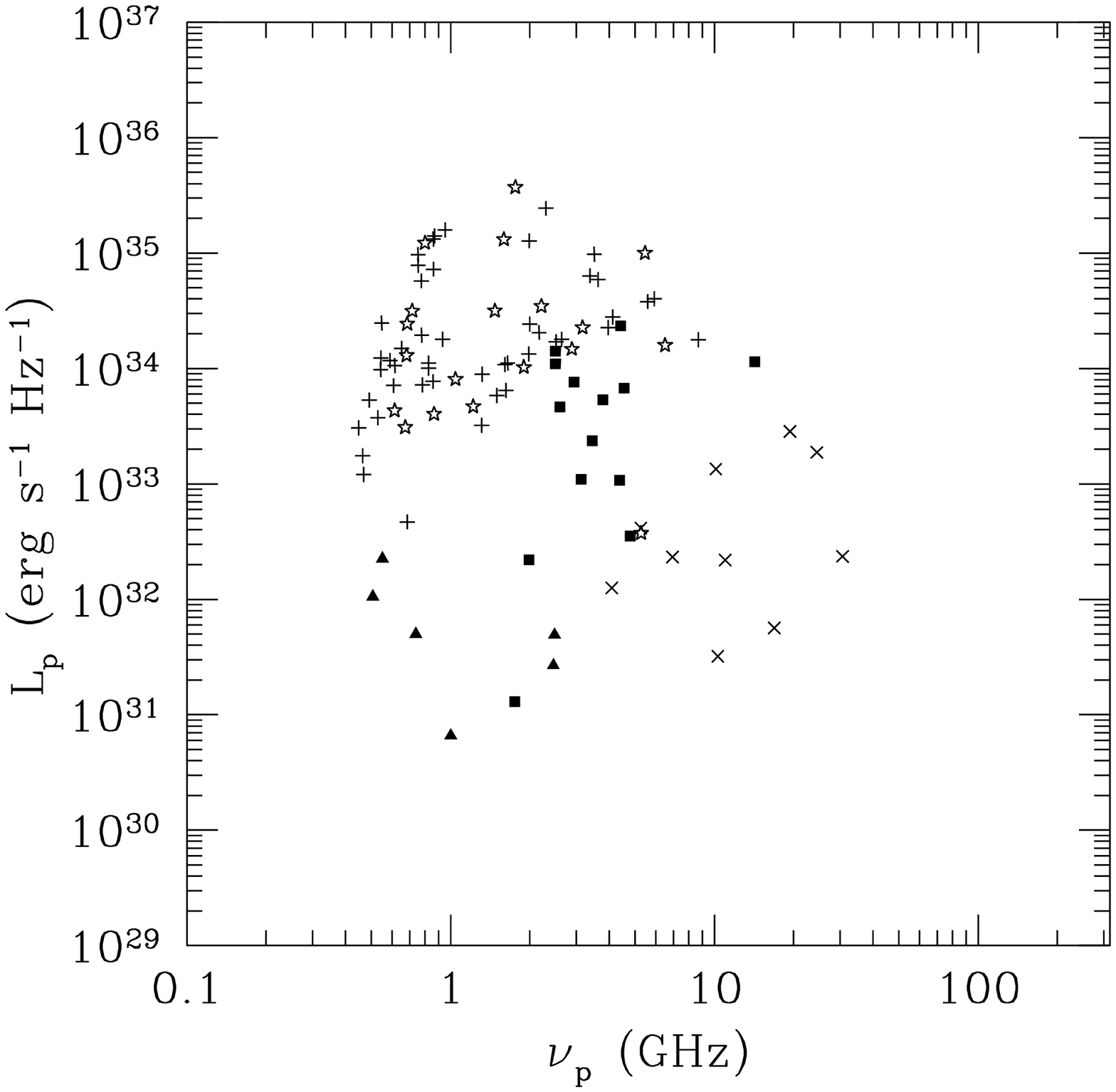,height=7cm} }}
\caption{\label{nup_Sp_Lp} Peak flux density versus observed peak
frequency (left) and peak luminosity versus rest frame turnover
frequency (right) for GPS galaxies in the various samples. HFP:
filled circles; Bolton sample A: $\times$; WENSS: filled squares;
CORALZ: filled triangles; Stanghellini: $\star$; Parkes: $+$.}
\end{figure*}

\subsection{The WMAP survey}

The WMAP point source catalogue (Bennett et al. 2003b) comprises
208 point sources with a $\ge 5\,\sigma$ detection in any of the 5
WMAP bands, in the range 22.8--93.5 GHz; 203 sources have
counterparts in existing 4.85 GHz, the remaining 5 sources being
probably spurious.

This is the first simultaneous multifrequency survey at mm
wavelengths and is therefore well suited to select extreme HFPs.
Although the completeness limit is at $S_{22.8\rm{GHz}} \simeq
1.25\,$Jy (De Zotti et al. 2005), to avoid that the estimates of
spectral indices are too affected by measurement errors we have
confined ourselves to sources with $S_{22.8\rm{GHz}} \ge 2\,$Jy,
and we have picked out those with inverted spectrum ($\alpha <0$)
from 22.8 to 33 GHz and from 4.85 to 22.8 GHz. There are 18
sources satisfying these criteria, a number close to the
prediction of the De Zotti et al. (2000) model, for a maximum
initial peak frequency of 200 GHz. Most of these sources are well
known calibrators and have therefore many observations at many
frequencies (see Trushkin 2003). They all show strong variability,
consistent with that observed for blazars (see Tinti 2005); 9 of
them are classified as blazars by Terasranta et al. (1998) or
Donato et al. (2001). The data on the WMAP sample thus confirm
that most quasars showing a peak at tens of GHz are likely blazars
caught during a phase when a flaring, strongly self-absorbed
synchrotron component dominates the emission spectrum (Tinti et
al. 2005).

\begin{figure}
\center{{ \epsfig{figure=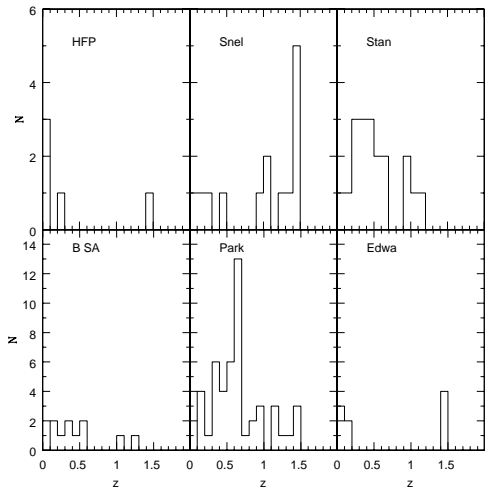,height=7cm} }}
\caption{\label{dis_z} Distributions of measured and estimated
redshifts for the HFP, Stanghellini et al.\ (1998), Snellen et
al.\ (1998), Bolton et al.\ (2004, sample A), Parkes, and ATCA GPS
galaxy samples. Note that the peaks at $z=1.5$ are due to
unidentified sources, assumed to be galaxies at that redshift.}
\end{figure}

\section{Impact of different selection criteria
\label{sect:selection}}

The observational properties that generally define the different
samples are:
\begin{itemize}
\item the flux density at the frequency of selection;
\item the turnover frequency range;
\item the spectral index in the optically thin and thick parts of the
spectrum.
\end{itemize}

\noindent To effectively explore the luminosity and peak frequency
evolution of GPS sources, we need samples that provide a wide
coverage of the $\nu_p$--$S_p$ plane. The coverage provided by the
present samples is shown in Fig.~\ref{nup_Sp_Lp}. HFP sources
(Dallacasa et al. 2000) have turnover frequencies greater than 4.9
GHz, at or above the upper limits of the samples of Stanghellini
et al. (1998) and Snellen et al. (1998, 2002, 2004). Because of
their high selection frequency, 15 GHz, the Bolton et al. (2004)
samples explore the high peak frequency region.

The samples by Bolton et al. (2004) have flux limits similar to
the faint GPS sample selected by Snellen et al. (1998), although
at a very different frequency, while the HFP and the Stanghellini
et al. (1998), and the Parkes samples contain bright sources.

The redshift distributions of the galaxies in the various samples
are shown in Fig.~\ref{dis_z}. Many redshifts are estimated from
optical magnitudes. The peak at $z=1.5$ is mostly due to optically
unidentified GPS sources, assumed, following Snellen et al.
(2000), to have that redshift.

\begin{figure*}
\center{{\epsfig{figure=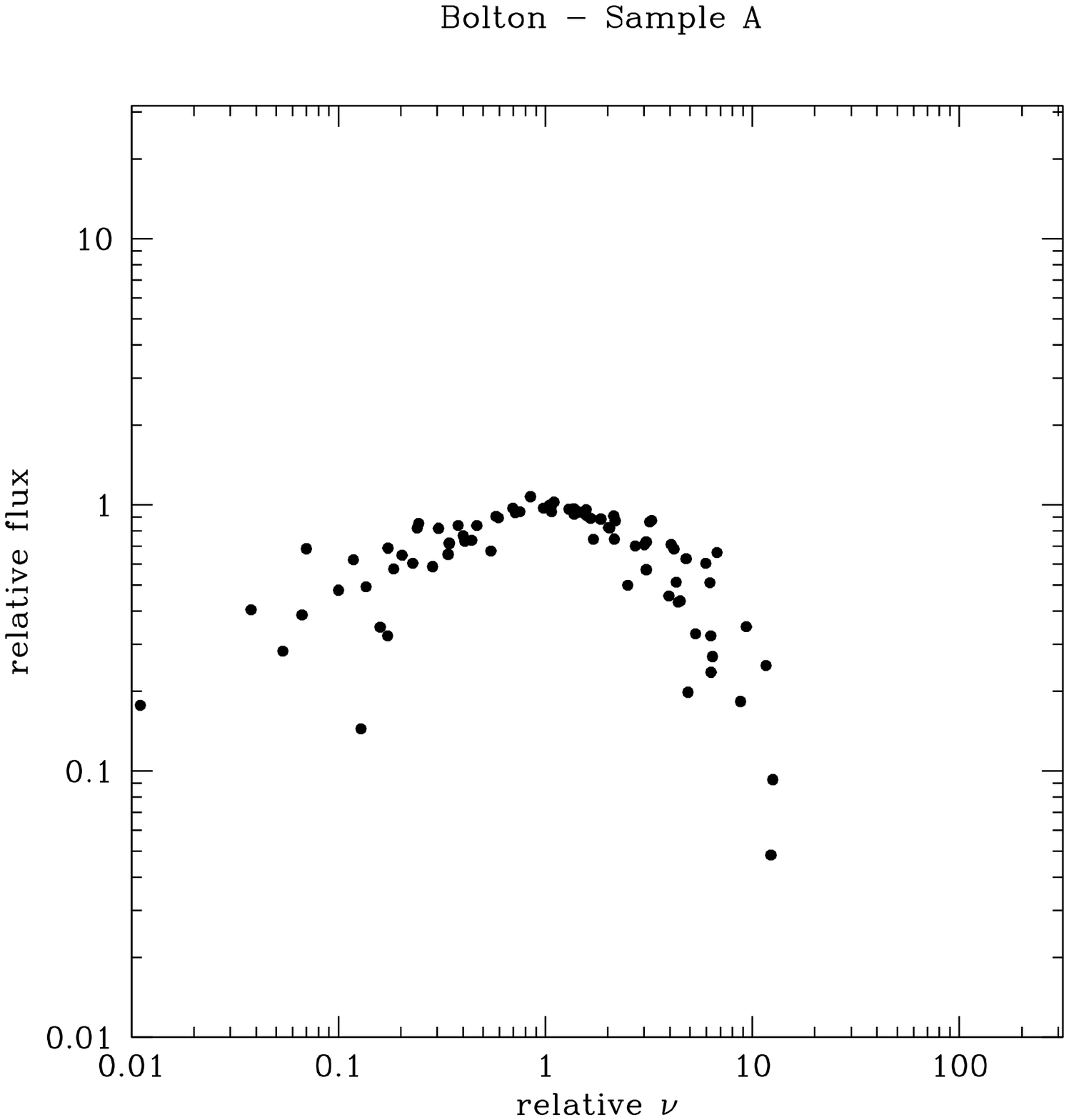,height=5cm}
\epsfig{figure=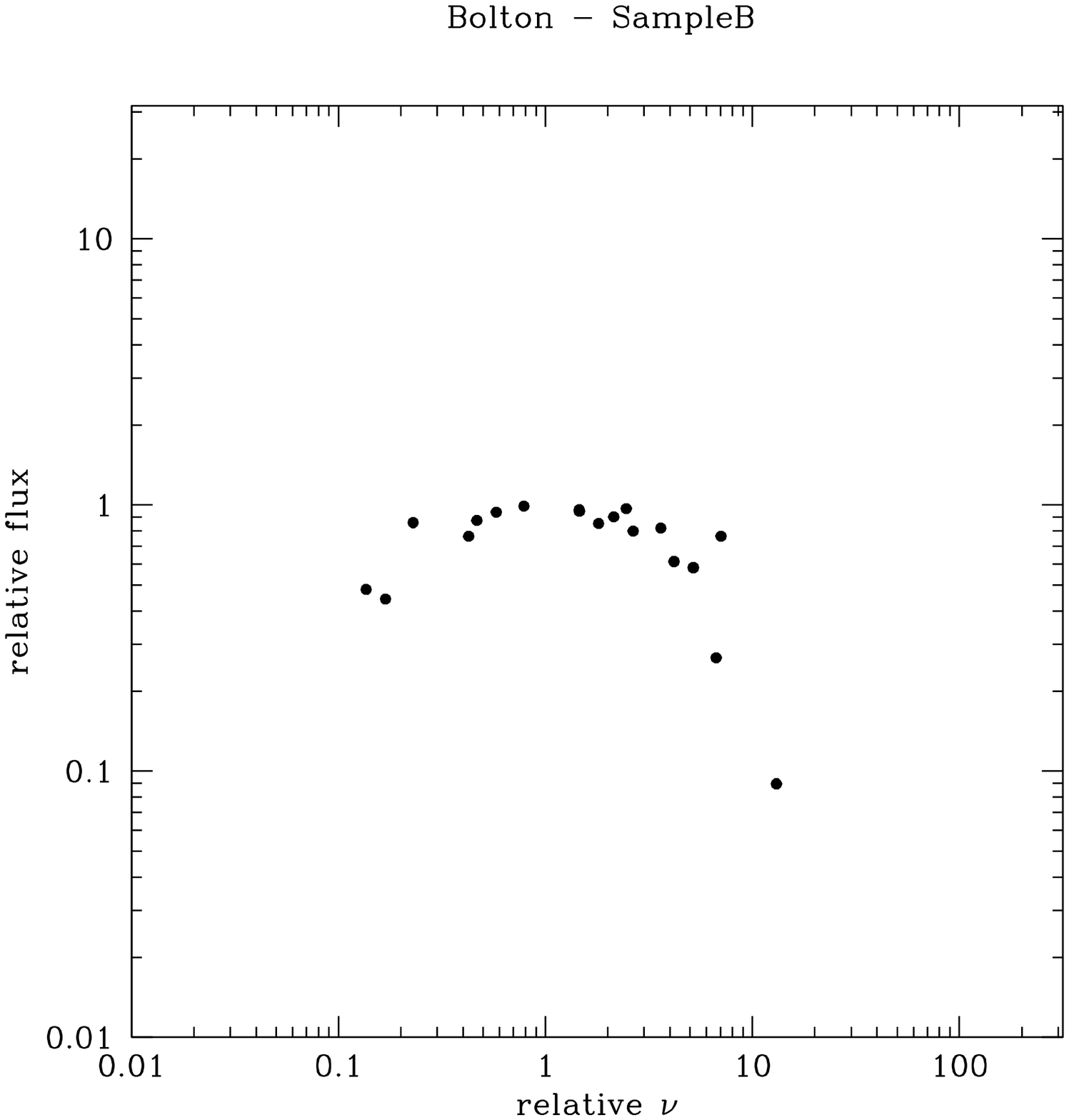,height=5cm}
\epsfig{figure=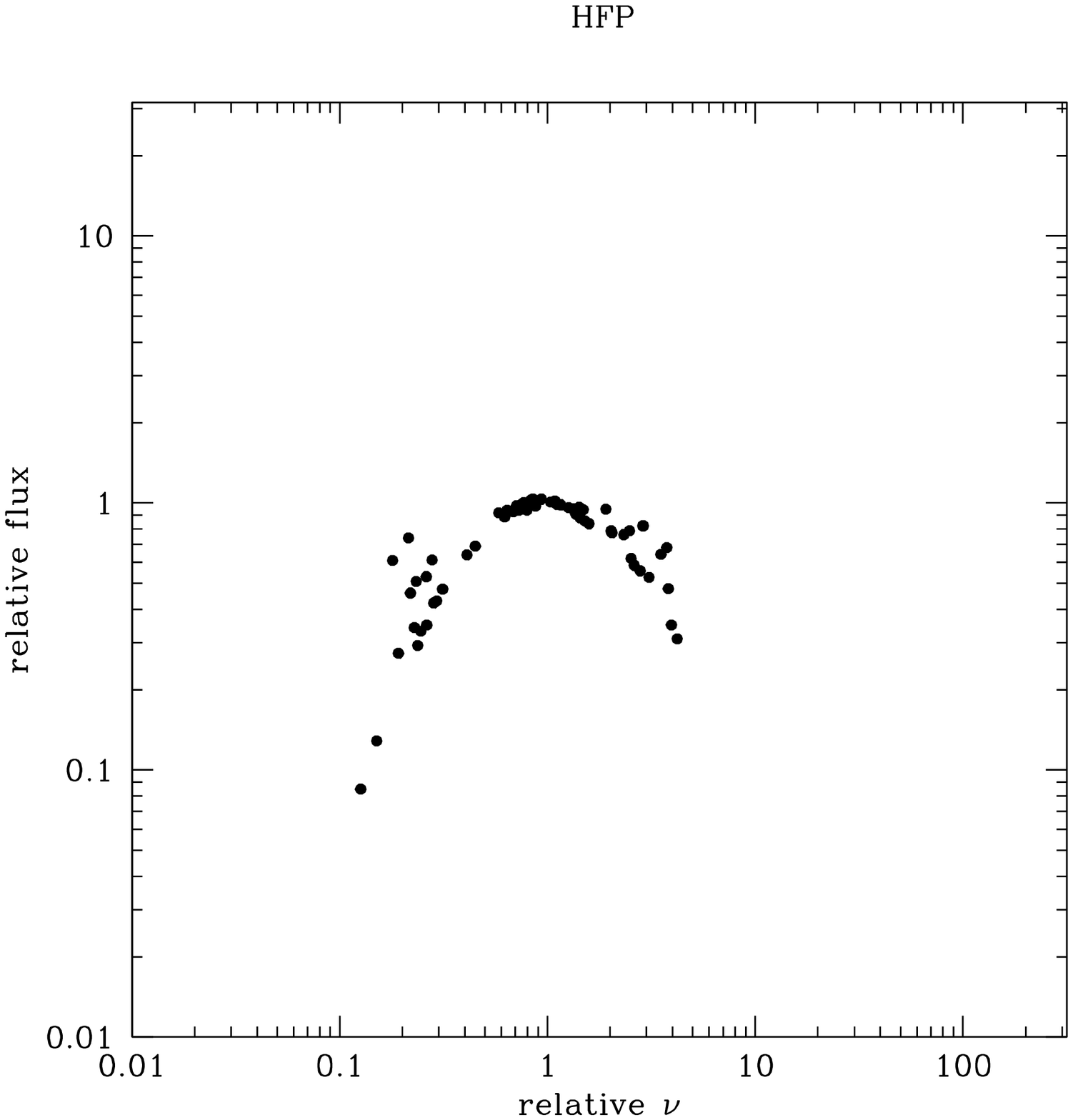,height=5cm}
\epsfig{figure=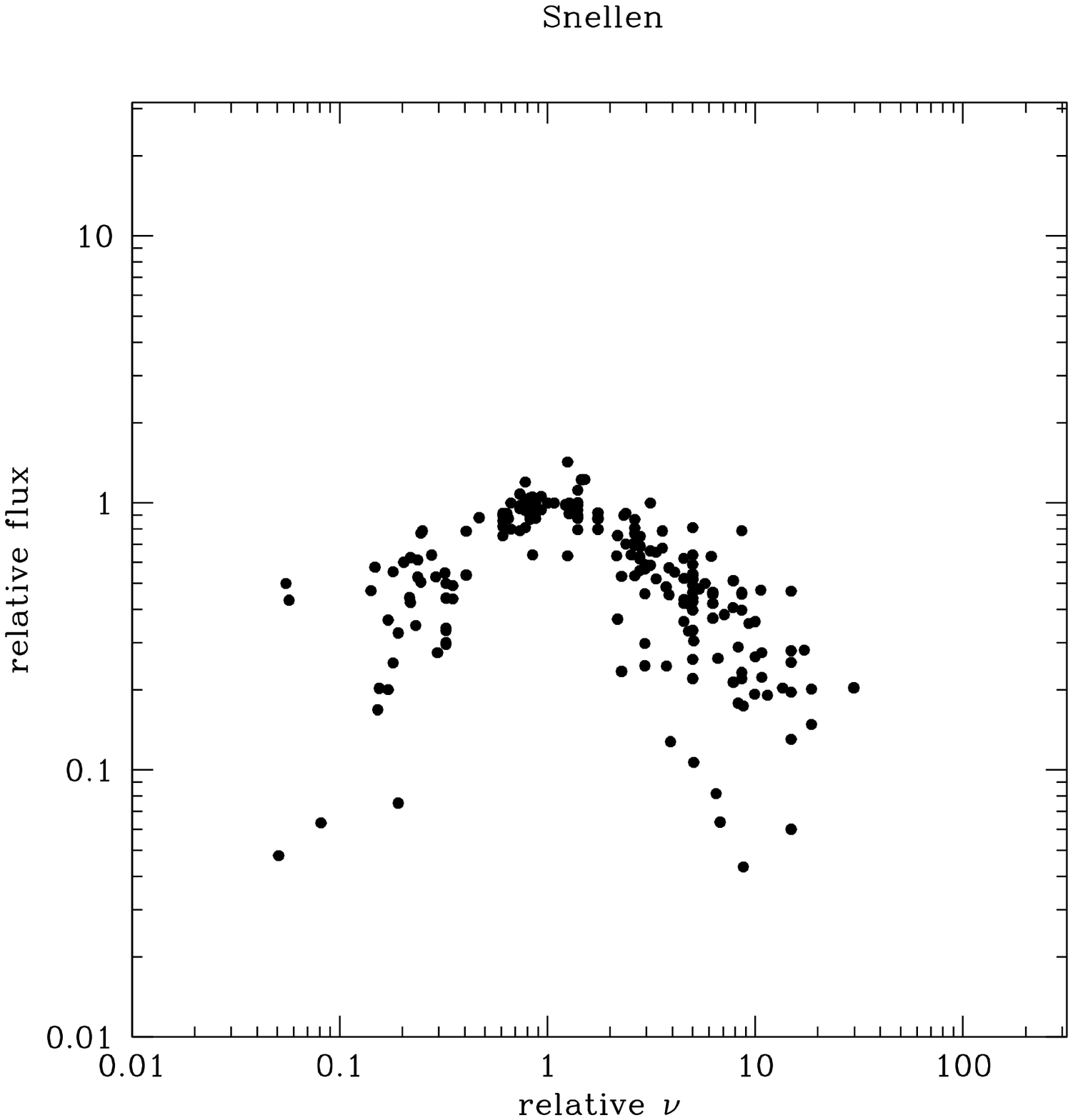,height=5cm}
\epsfig{figure=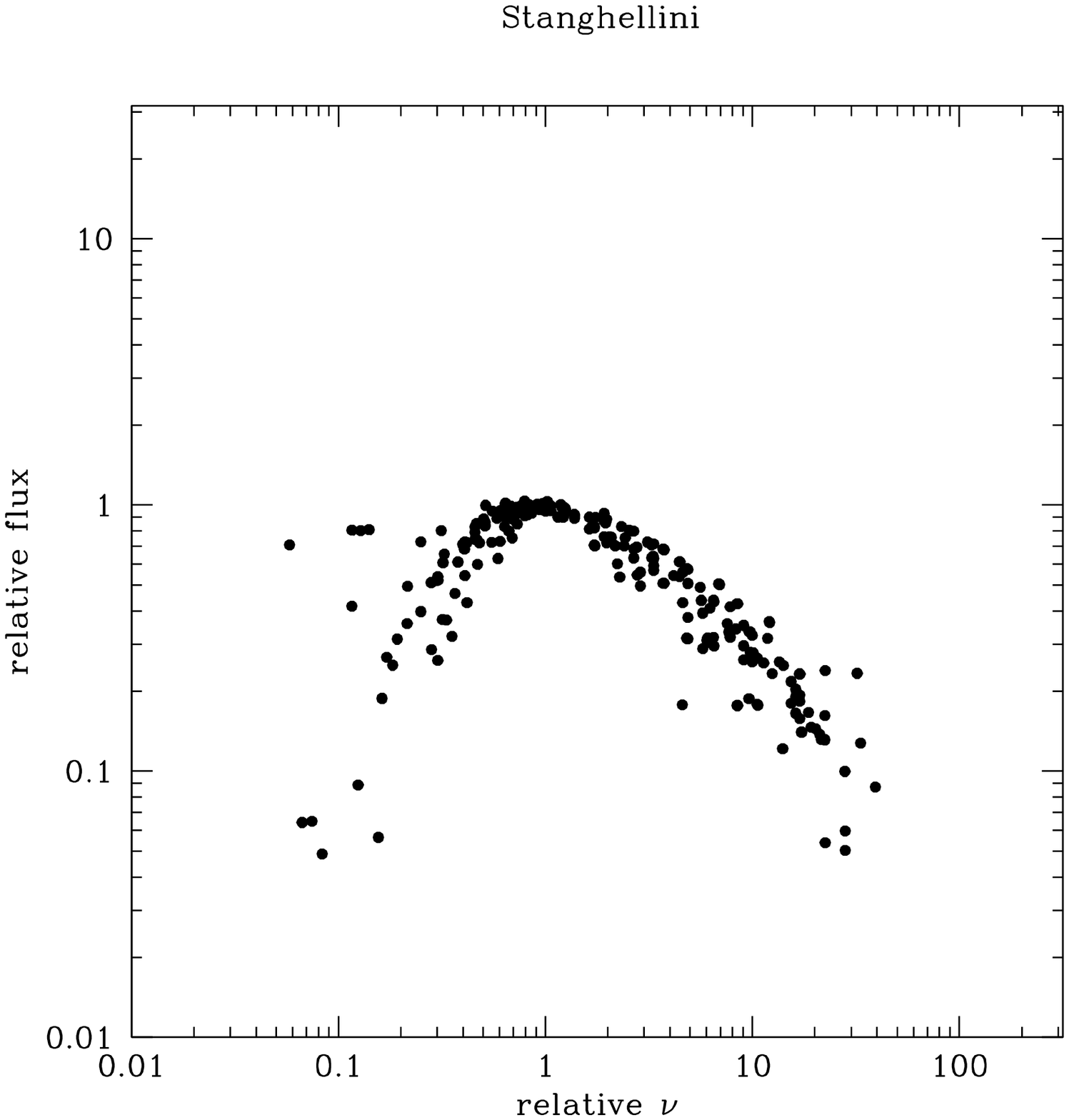,height=5cm}
 }}
\caption{\label{spettri_compositi} Spectra of GPS sources in the
various samples, normalized in both frequency and flux density. }
\end{figure*}

\begin{table}
\begin{center}
\caption{\label{spc_fits} Median values of the optically thick and
thin spectral indices for the various samples.}
\begin{tabular}{lll}
\hline\hline
Sample     & $\alpha_{\rm thick}$ & $\alpha_{\rm thin}$\\
\hline
HFP    & -0.86 & 0.61 \\
Stan   & -0.76 & 0.67 \\
B\_SA  & -0.38 & 0.76 \\
B\_SB  & -0.43 & 0.82 \\
Snel   & -0.67 & 0.64  \\
\hline
\end{tabular}
\end{center}
\end{table}

As noted above, selection criteria include constraints on the
spectral indices in the optically thick and/or in the optically
thin spectral region. Stanghellini et al. (1998) required
$\alpha_{\rm thin} \ge 0.5$, Dallacasa et al. (2000) demanded
$\alpha_{\rm thick} \le -0.5$; Edwards \& Tingay (2004) applied
both constraints. For other samples, the adopted spectral criteria
are less explicit. Snellen et al. (1998, 2000) require an inverted
spectrum at low frequencies and that the Full Width at Half
Maximum of the spectrum is less than 2 decades in frequency; the
latter condition implies that, typically, $\alpha_{\rm thin}$ and
$-\alpha_{\rm thick}$ are $\ge 0.3$, a somewhat less restrictive
constraint than adopted for the previously mentioned samples. An
even looser criterion was adopted by Bolton et al. (2004):
$\alpha^{4.8}_{1.4} < -0.1$. Not many details are given on
spectral criteria for the Parkes sample (Snellen et al. 2002),
while the CORALZ sample (Snellen et al. 2004) was selected on the
basis of radio morphology, not of the spectra.

In Fig. \ref{spettri_compositi} we compare the shapes of the radio
spectra of the sources in the different samples, normalized to the
peak frequencies and to the peak flux densities. It is apparent,
and quantified in Table~\ref{spc_fits}, that the median
$\alpha_{\rm thick}$ of Bolton et al. (2004) sources is
considerably flatter than for the other samples. To homogenize
this sample to the others, we have dropped sources with
$\alpha_{\rm thick}> -0.5$.

\begin{table}
\caption{Best fit values of the model parameters for GPS
galaxies.} \label{parameters}
\begin{center}
\begin{tabular}{lr}
\hline\hline $\log(n_0)\,[\hbox{Mpc}^{-3}\,(d\log
L_\nu)^{-1}\,\hbox{GHz}^{-1}]$ & -14.13 \\
$\beta$                         & 0.83   \\
$\log(\nu_{p,i})\,$ (GHz)       & 2.98 \\
$\eta$                          & 1.20 \\
$\lambda$                       & 0.0125 \\ \hline
\end{tabular}
\end{center}
\end{table}

\begin{figure*}
\center{{
\epsfig{figure=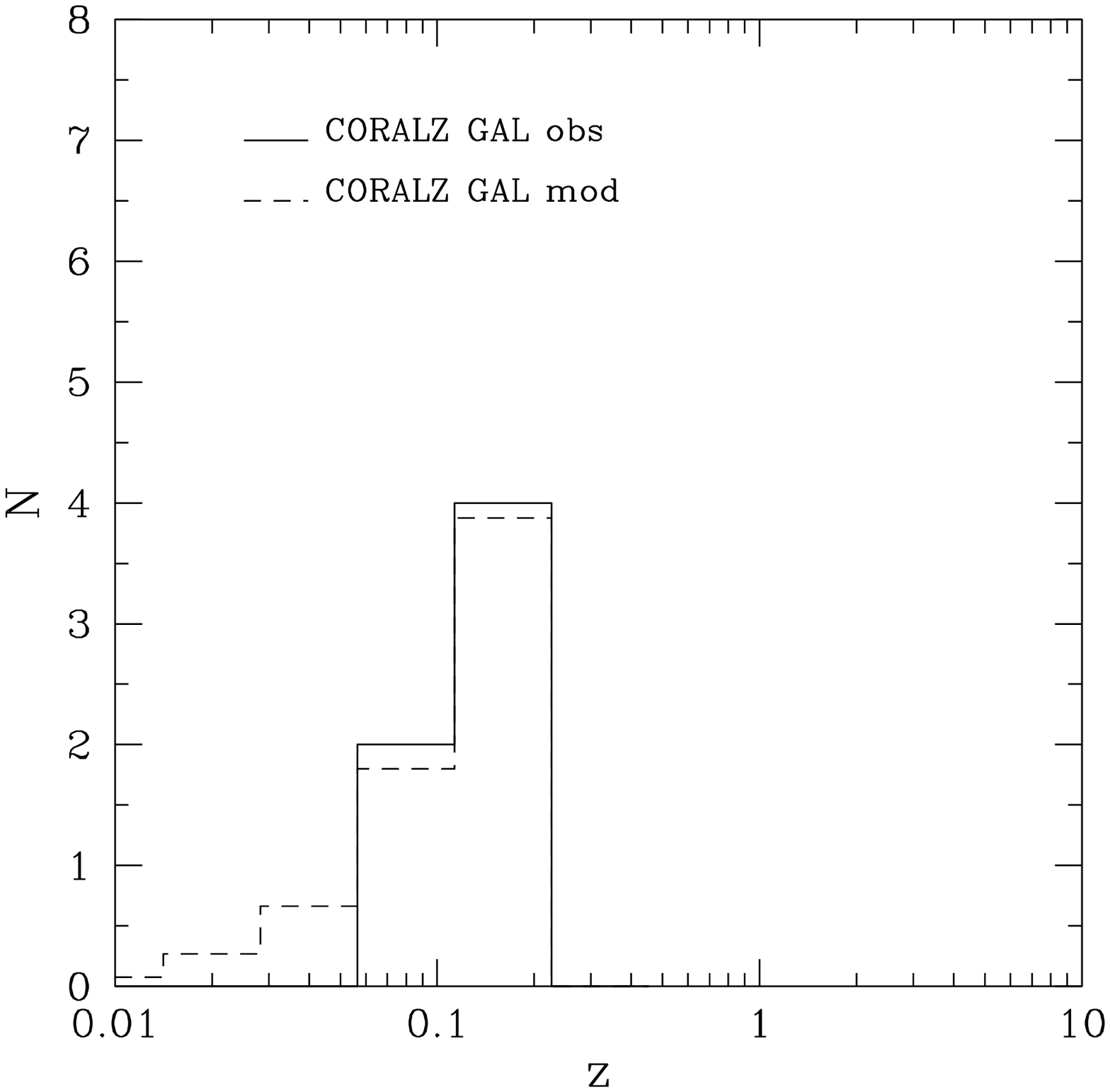,height=5cm}
\epsfig{figure=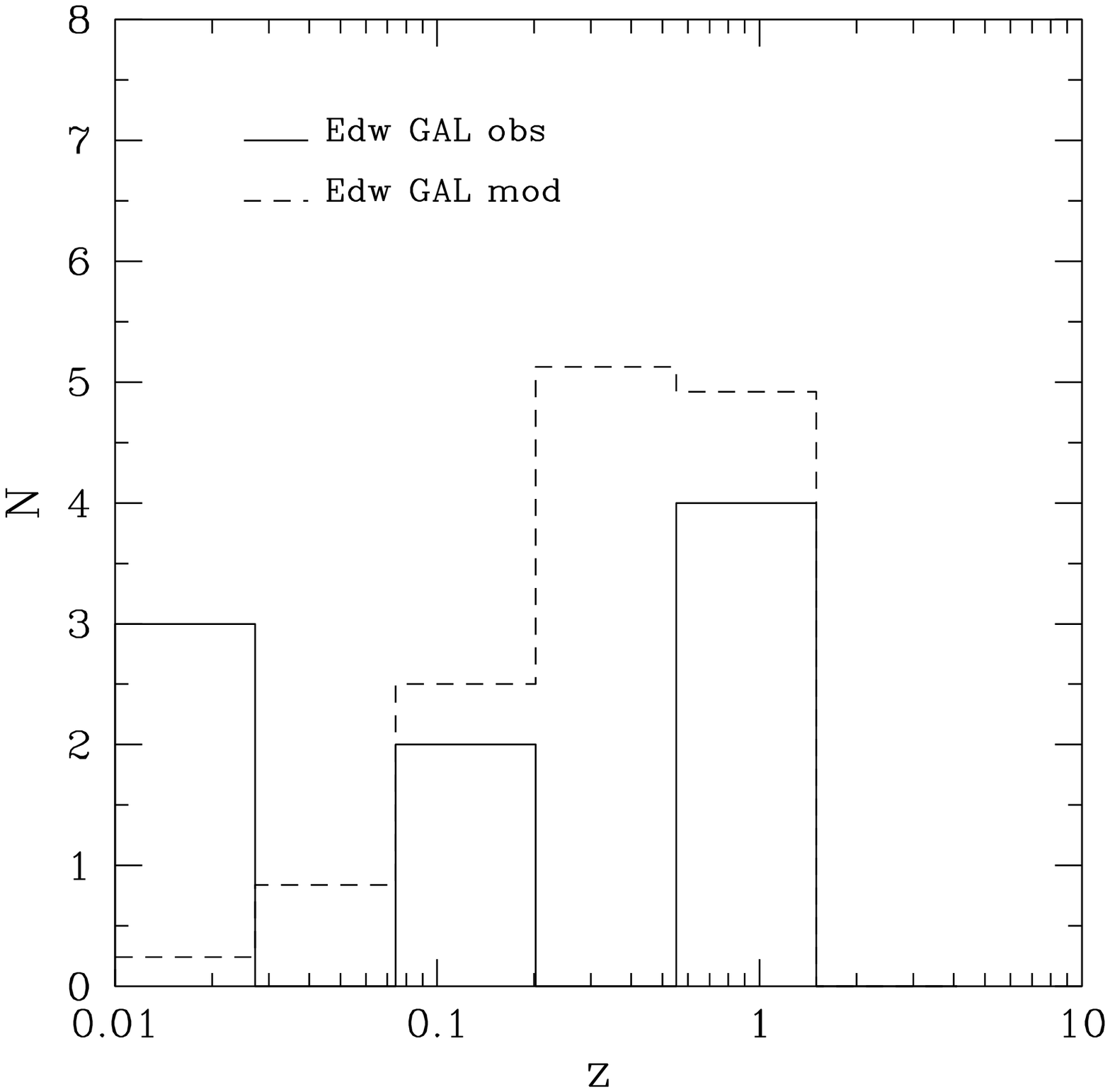,height=5cm}
\epsfig{figure=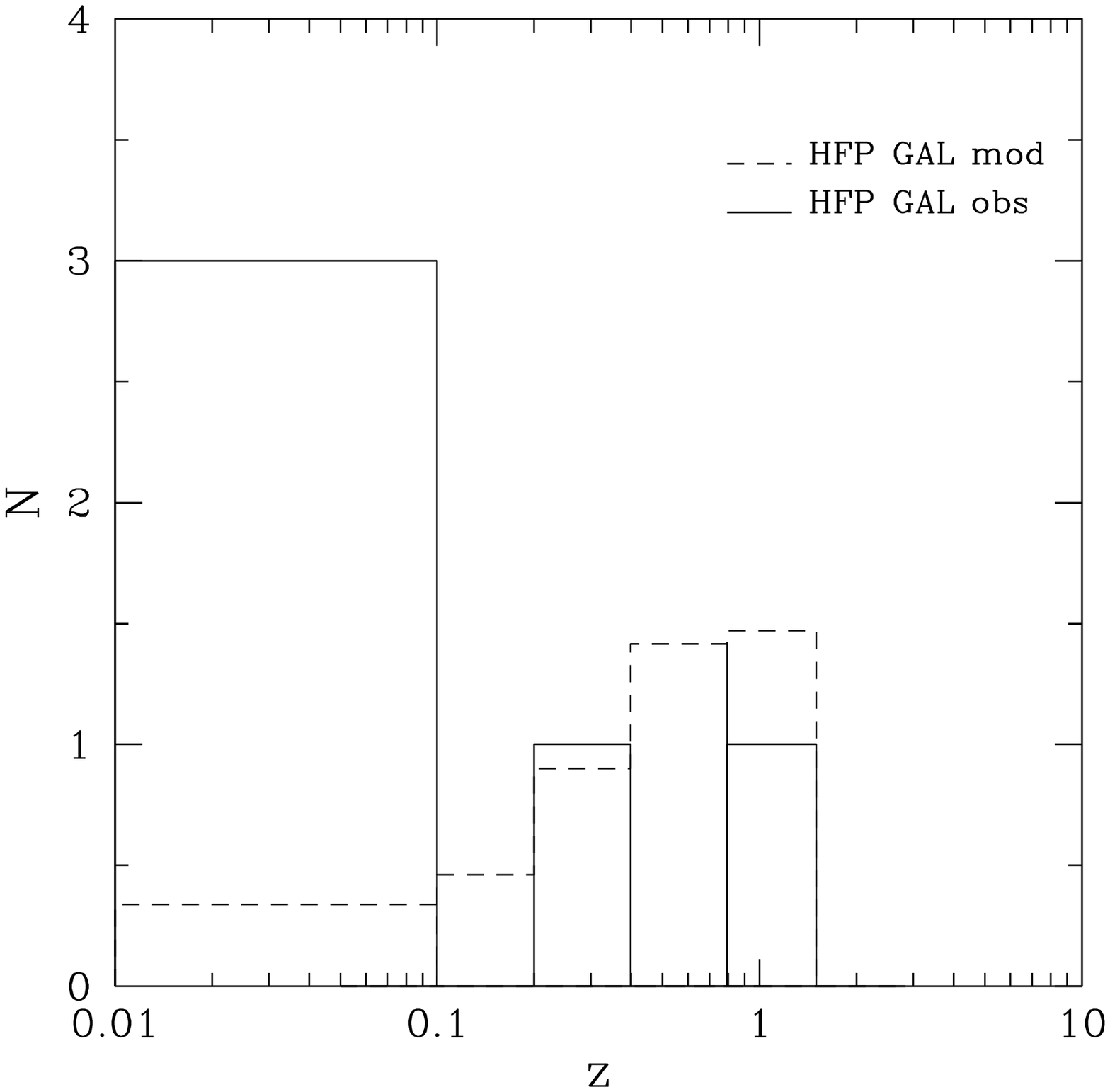,height=5cm}
\epsfig{figure=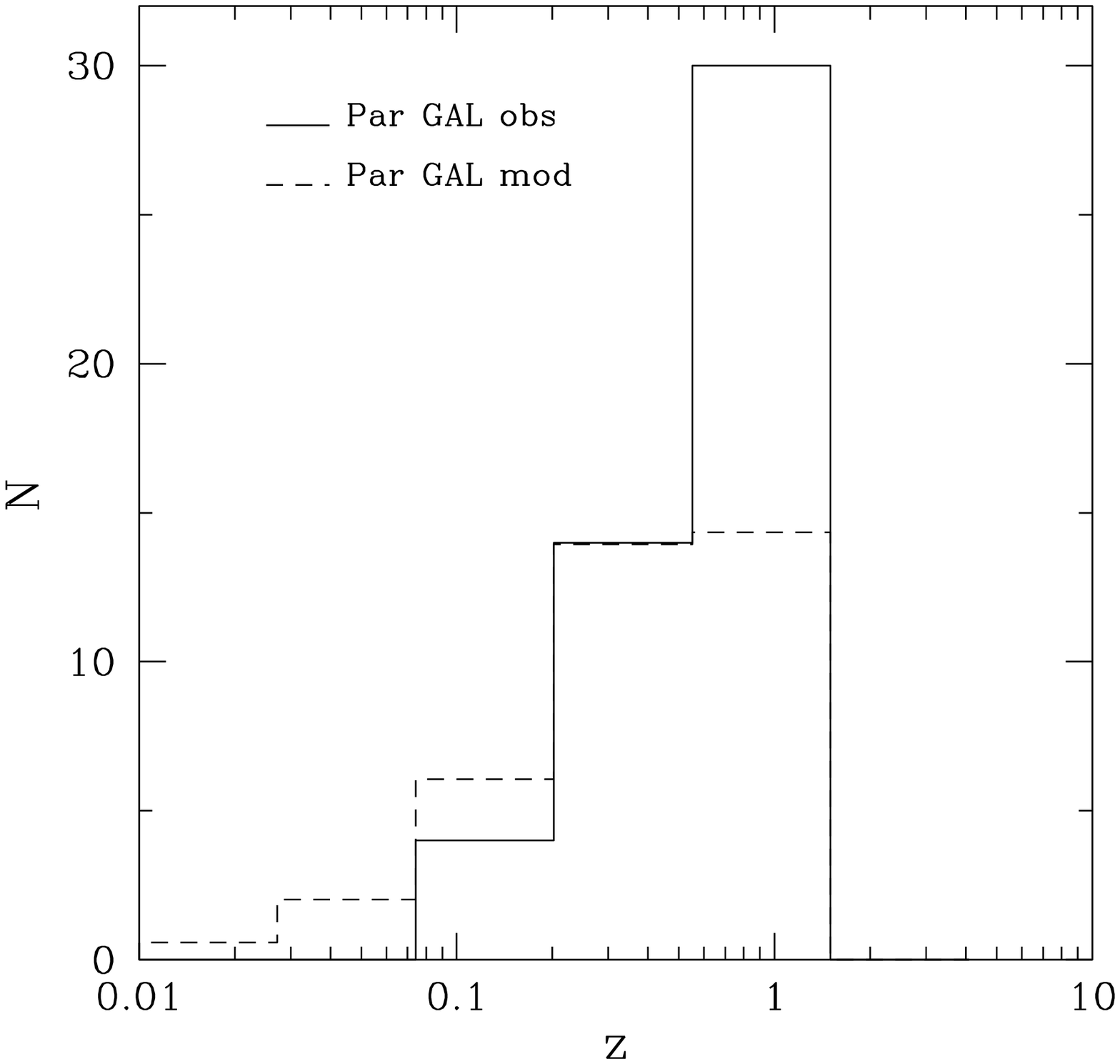,height=5cm}
\epsfig{figure=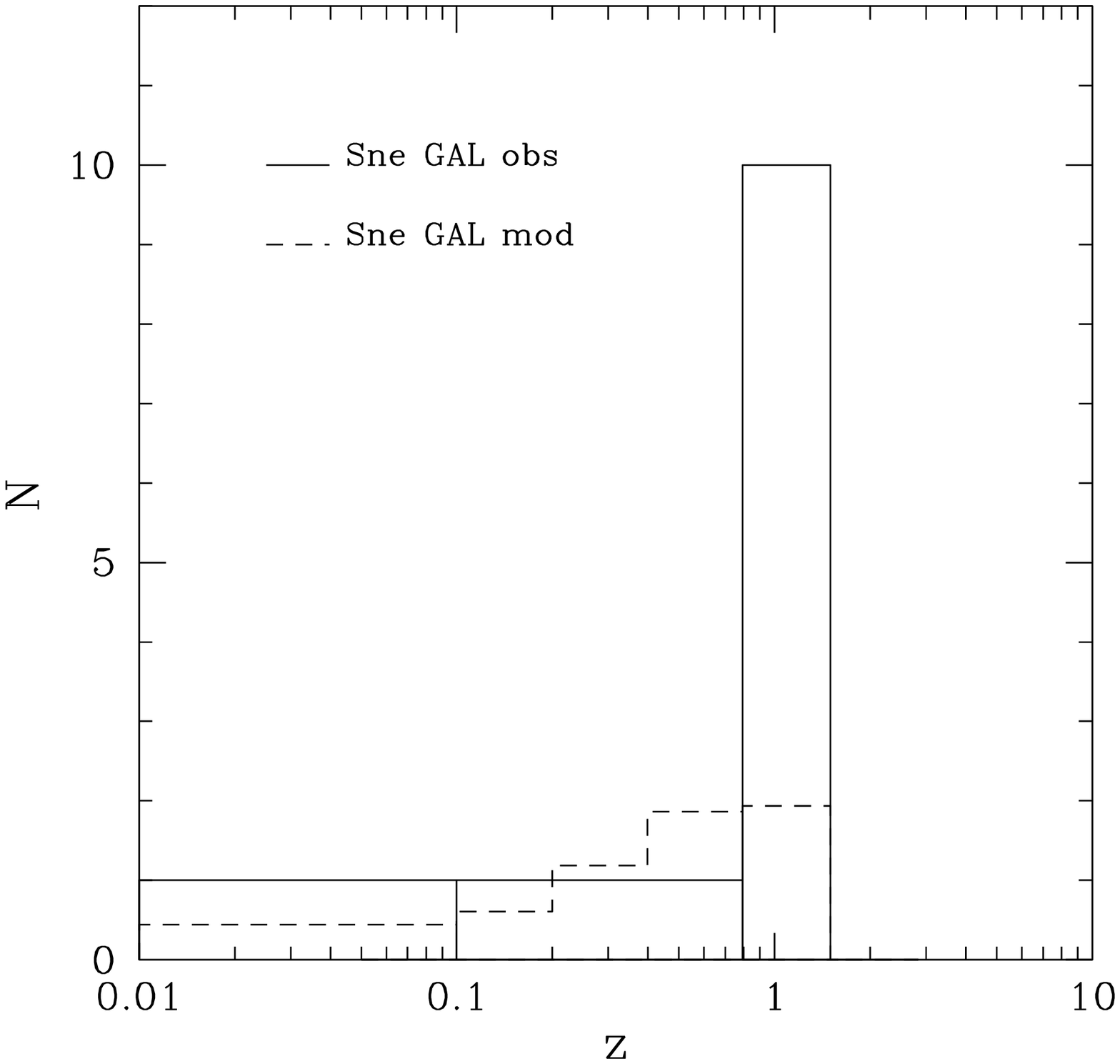,height=5cm}
\epsfig{figure=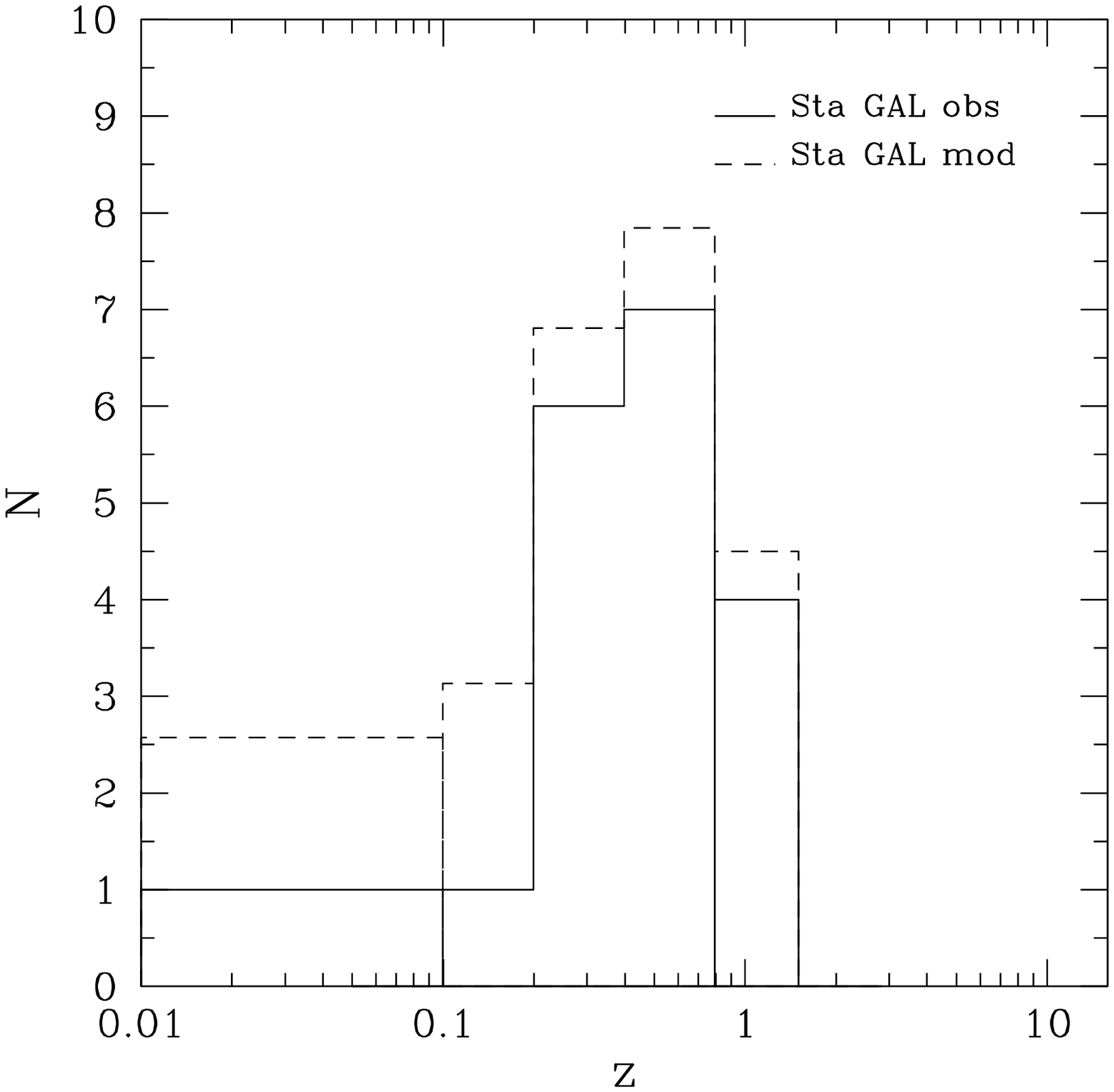,height=5cm}
 }}
\caption{\label{distr_z} Comparison of the redshift distributions
yielded by the best fit model (dashed) with the observed ones
(solid). }
\end{figure*}

\begin{figure*}
\center{{
\epsfig{figure=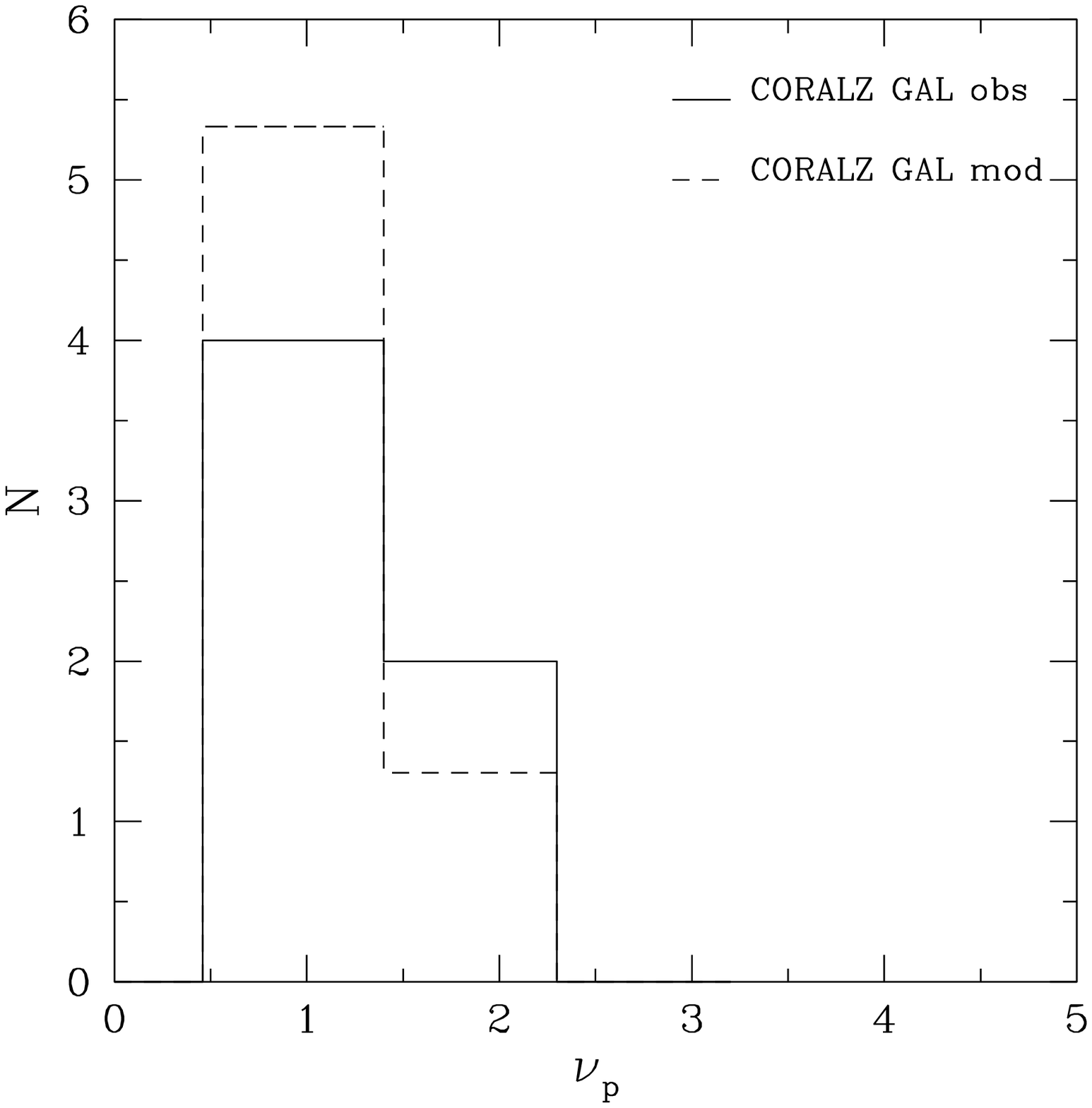,height=5cm}
\epsfig{figure=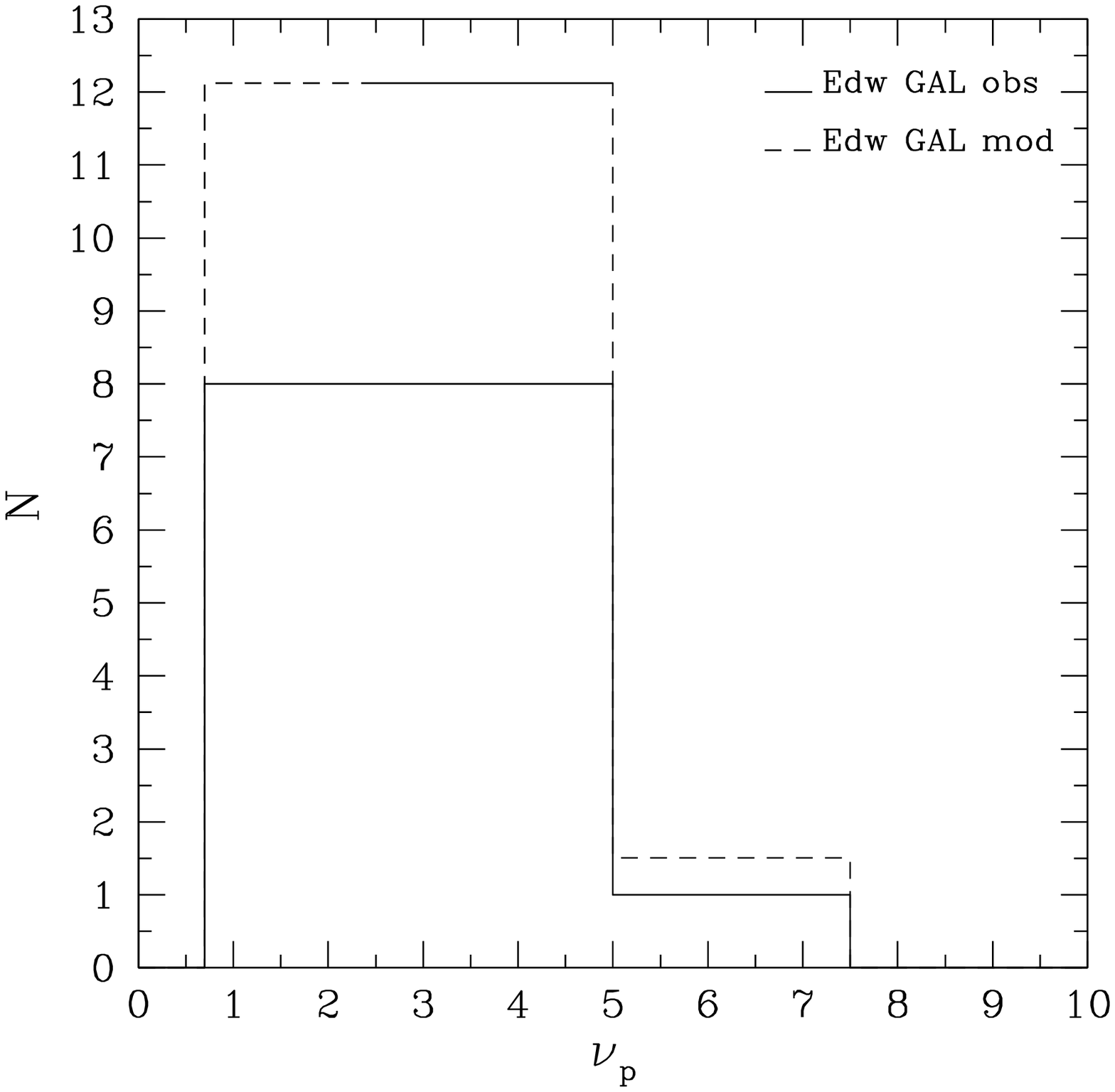,height=5cm}
\epsfig{figure=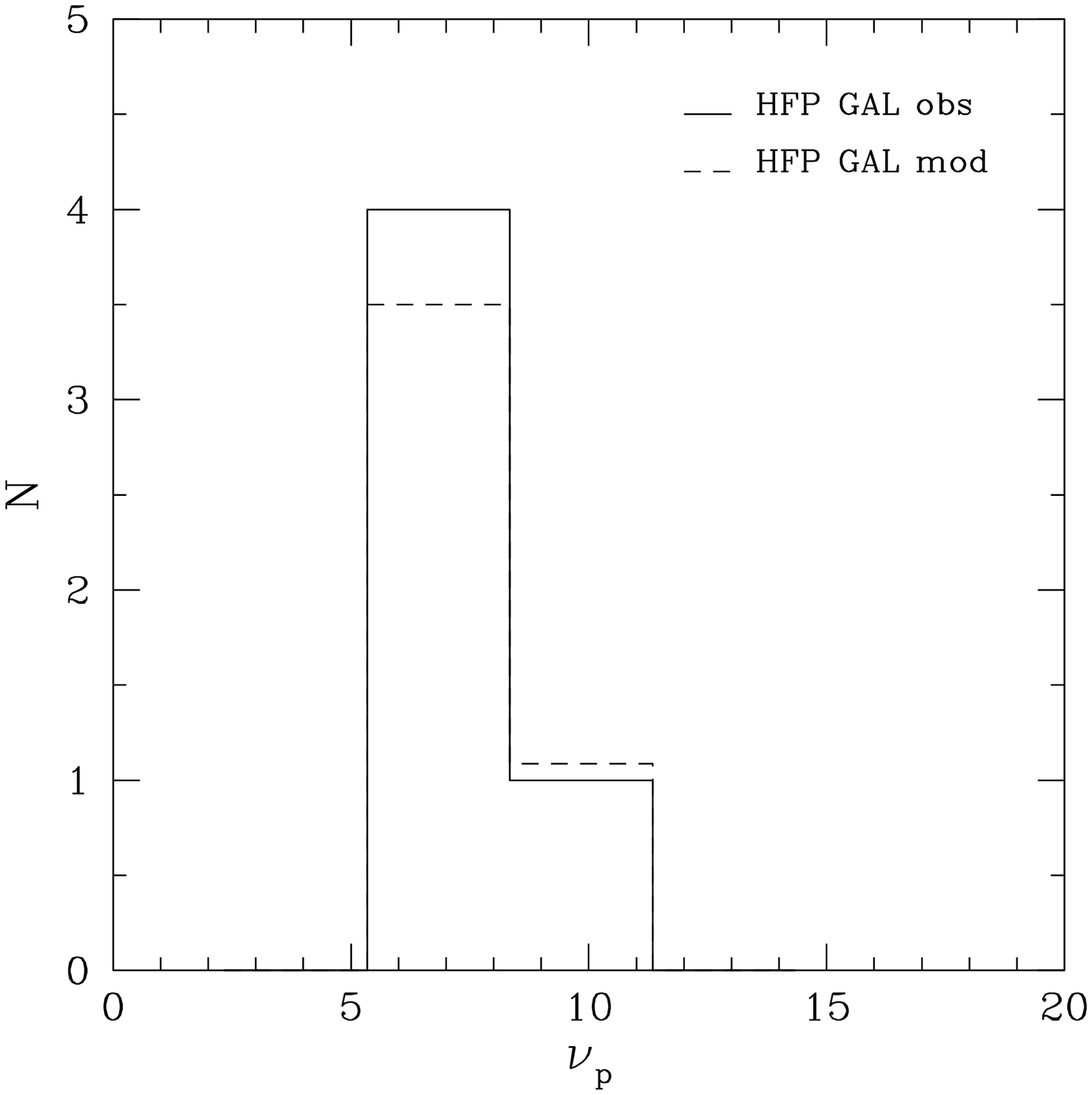,height=5cm}
\epsfig{figure=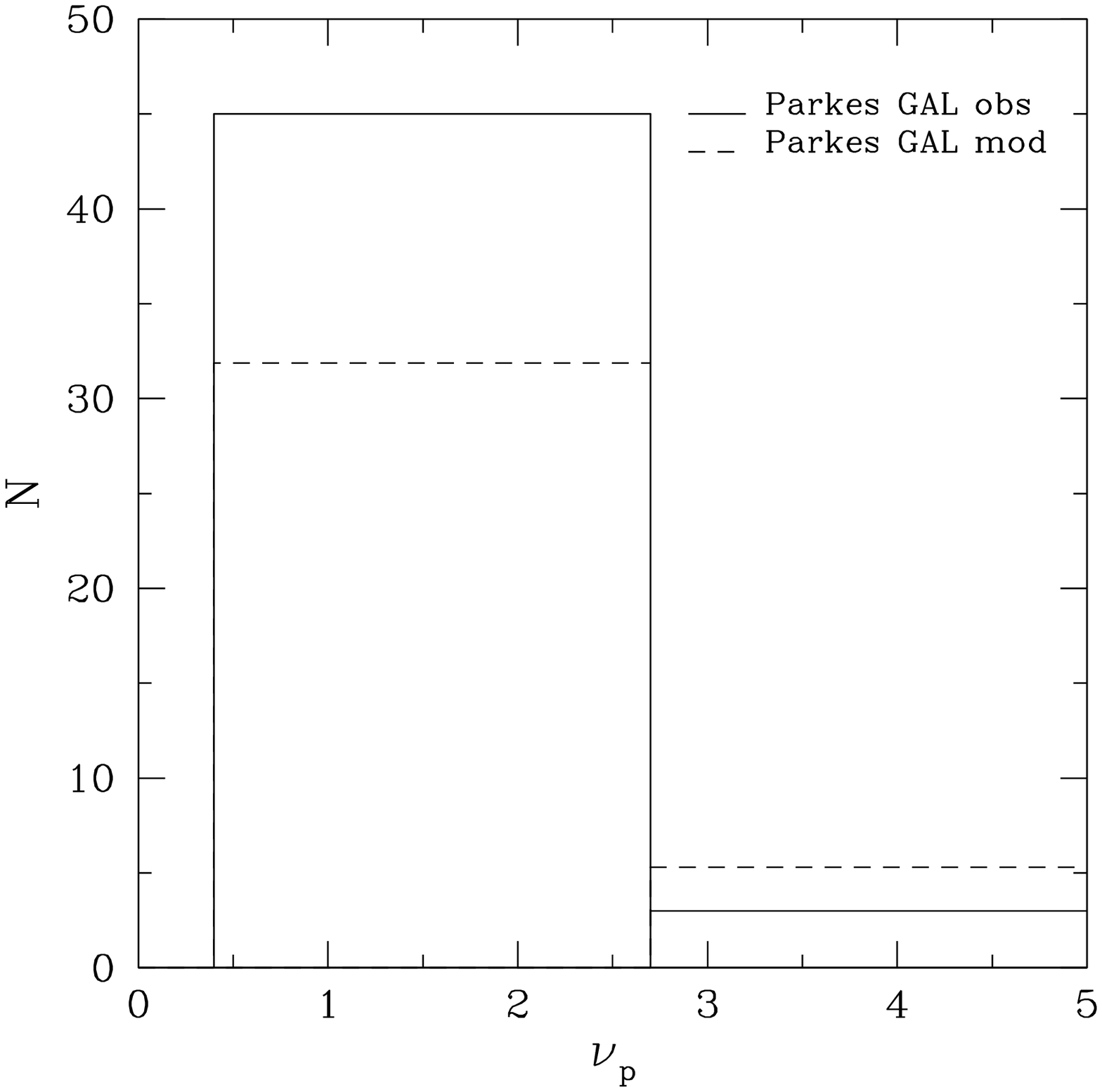,height=5cm}
\epsfig{figure=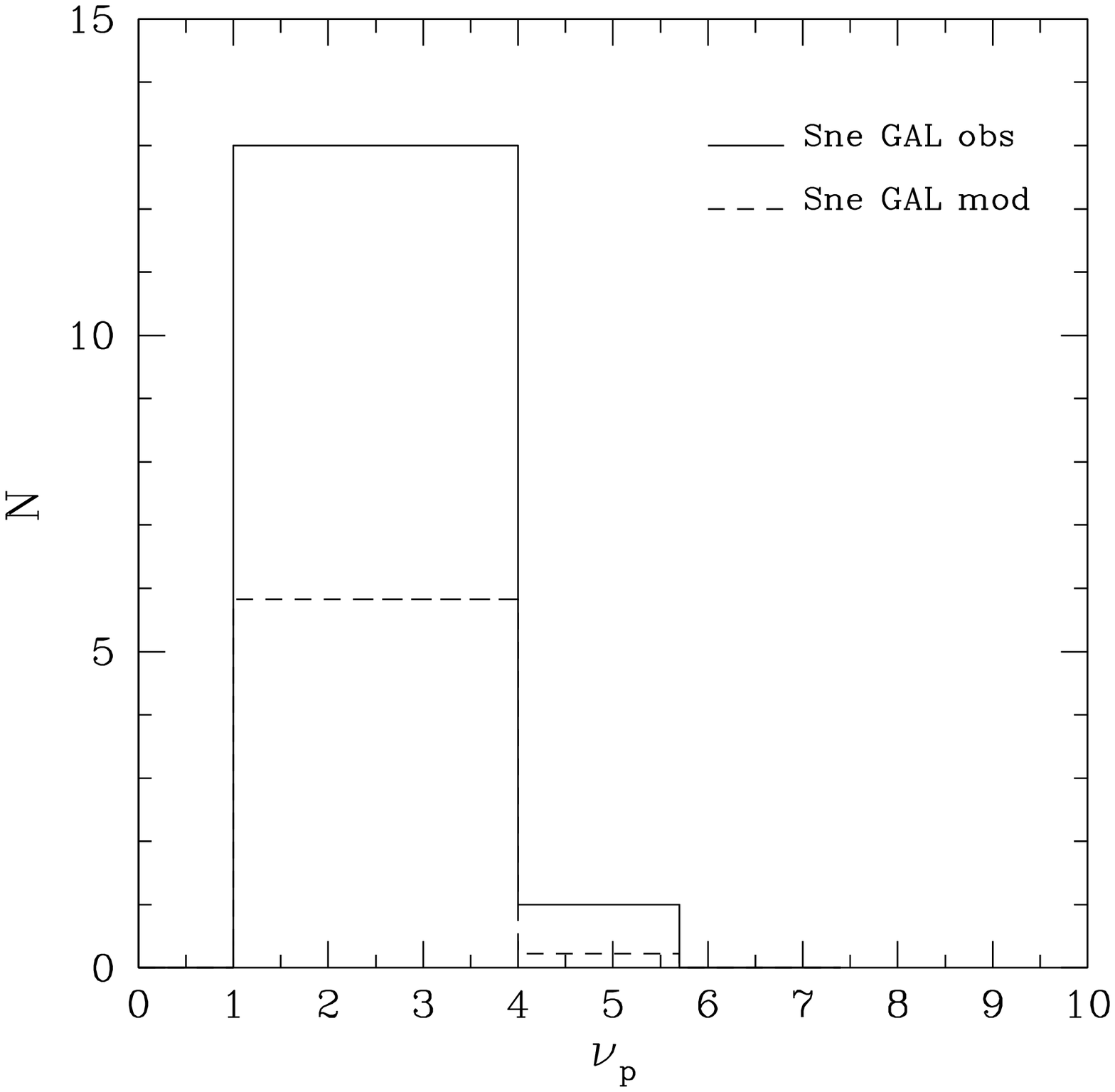,height=5cm}
\epsfig{figure=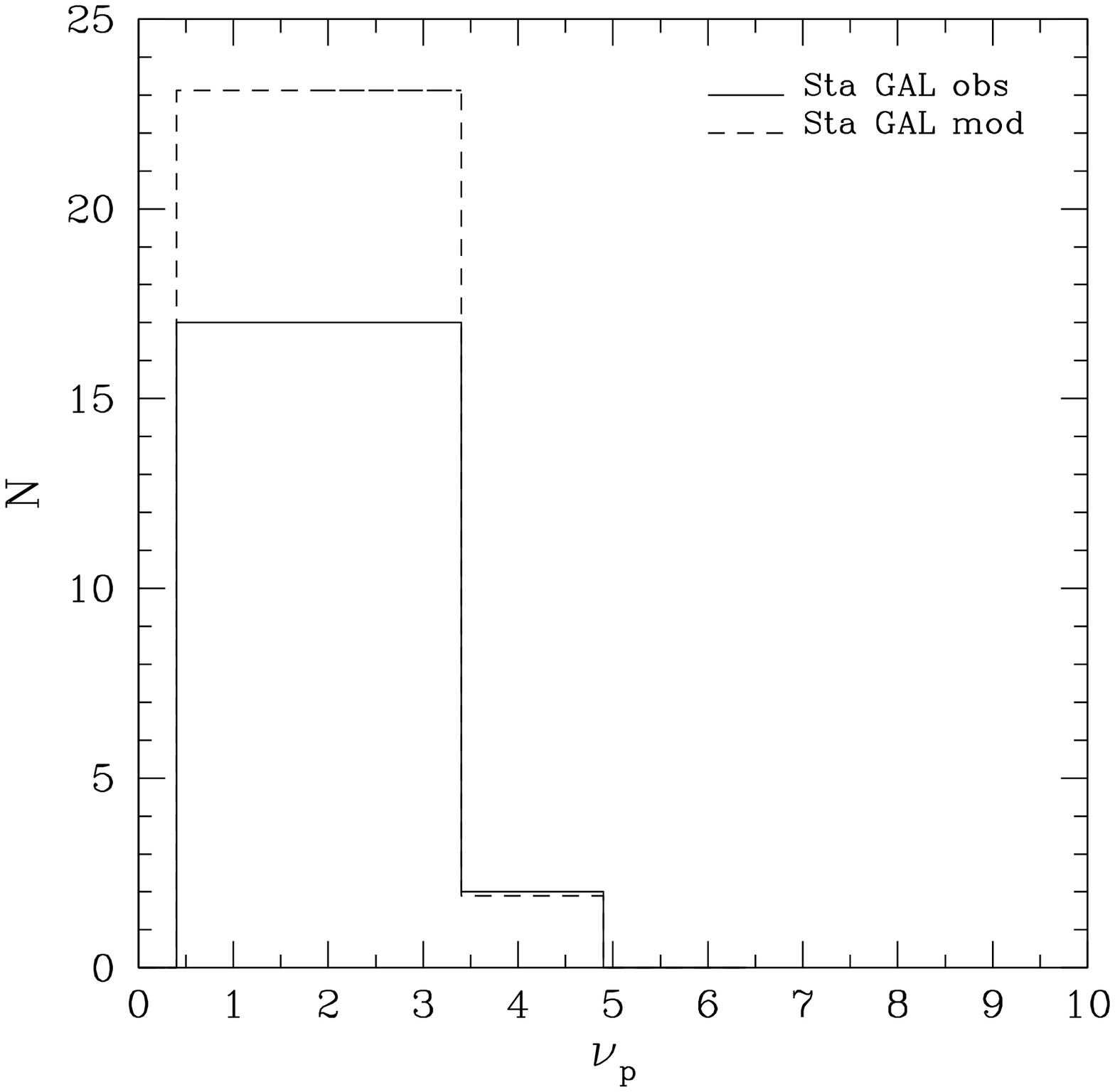,height=5cm}
 }}
\caption{\label{distr_nup} Comparison of the peak frequency
distributions yielded by the best fit model (dashed) with the
observed ones (solid). }
\end{figure*}

\section{Results}

We have fitted separately the redshift and the observed peak
frequency, $\nu_{p,0}$, distributions of GPS galaxies in the
samples described above. The best fit values of the parameters
have been obtained minimizing the chi-square function with the
Minuit package (CERN libraries). We need to determine:
\begin{itemize}
\item the parameters characterizing the luminosity function, i.e.
the normalization $n_0$ and the slope $\beta$; the minimum and
maximum luminosities, ${\rm L_{\rm min}}$ and ${\rm L_{\rm max}}$
respectively, both referred to the frequency of 300 MHz, below the
conventional minimum peak frequency of GPS sources, have been set
at the minimum and maximum observed values: $L_{\rm
max}(300\hbox{MHz}) = 5\cdot 10^{35}\,\hbox{erg}
\,\hbox{s}^{-1}\,\hbox{Hz}^{-1}$, $L_{\rm min}(300\hbox{MHz}) =
10^{29}\,\hbox{erg} \,\hbox{s}^{-1}\,\hbox{Hz}^{-1}$;
\item the slope, $\eta$, of the power-law dependence of the emitted radio power
on source age;
\item the slope, $\lambda$, of the power-law dependence of the
peak frequency on source age;

\item the initial value of the peak frequency, $\nu_{p,i}$;
\item the redshift of formation of the first peaked spectrum sources,
$z_f$;
\item the parameter $k$ characterizing the cosmological evolution
of the luminosity function.
\end{itemize}
We have checked that the data do not require cosmological
evolution of the luminosity function of GPS galaxies, confirming
the finding of De Zotti et al. (2000), and we have therefore set
$k=0$. We thus have $L_\star(z)=\hbox{const}= L_0=
10^{32}\,\hbox{erg}\, \hbox{s}^{-1}\,\hbox{Hz}^{-1}$
[Eq.~(\ref{eq:evol})]. The fit is also insensitive to the value of
$z_f$, provided that it is larger than the maximum estimated
redshift of GPS galaxies in the considered samples. We have
therefore fixed $z_f=1.5$.

Figures~\ref{distr_z} and \ref{distr_nup} compare the model with
the observed redshift and peak frequency distributions. The
overall agreement is reasonably good, indicating that the
underlying scenario is consistent with the current data.

The best fit values of the parameters are given in
Table~\ref{parameters}. The formal errors on them derived from the
$\chi^2$ statistics are rather small, but we regard them as
unrealistic in view of the many uncertainties due to the
difficulties in the sample selection; also the redshift
distributions are largely built using photometric redshift
estimates rather than spectroscopic measurements. Thus we prefer
not to report uncertainties that are likely to be deceitful. That
the real uncertainties are probably large is indicated by a
comparison of the present best-fit values of the parameters with
those found by De Zotti et al. (2000) analyzing some of the
samples considered here. In particular, the values of $\eta$ and
$\lambda$ are widely different [the difference of $n_0$ follows
from that of $\lambda$, see Eq.~(\ref{eq:FL})]. Still both
analyses find positive values of $\eta$ and negative values of
$p=-\eta + \alpha\lambda$, implying that both the emitted radio
power and the peak luminosity {\it decrease} with increasing
source age, at variance with the evolution models by Snellen et al.
(2000) and Alexander (2000).

The slope of the luminosity function is more stable: we find
$\beta=0.83$ while De Zotti et al. (2000) found $\beta=0.75$. Both
values are not far from the slope of the low luminosity portion of
the luminosity function of steep spectrum radio sources,
$\beta_{ss} \simeq 0.69$ (Dunlop \& Peacock 1990; Magliocchetti et
al. 2002).

\section{Discussion and conclusions}

Although new samples have substantially improved the coverage of
the luminosity--redshift--peak frequency space of GPS galaxies,
the assessment of their evolutionary properties is still
difficult. First of all, there is no fully agreed set of criteria
to ascertain whether a source is truly a GPS. The frequently
adopted GPS identification conditions rely on the spectral shape.
However, as mentioned in Sects.~\ref{sect:samples} and
\ref{sect:selection}, different spectral requirements have been
adopted by different groups and recovering a homogeneous set of
data is not easy and may be even impossible. Other properties that
GPS sources should have, include a compact structure (size $\lsim
1\,$kpc), low polarization, low variability, and sub-luminal
component motions. Regrettably, measurements of these quantities
are available only for a limited number of GPS candidates. The
samples we are using may therefore be contaminated by sources of
different nature.

Also, a significant fraction of objects are still unidentified,
and spectroscopy of identified objects is highly incomplete, so
that many redshifts are estimated from optical magnitudes.
Although GPS galaxies seem to have a rather well defined
redshift--magnitude relationship, its dispersion is significant
and may increase with redshift.

With these premises we cannot expect to be able to come out with a
clear assessment of the evolutionary properties of GPS galaxies.
Still, some interesting conclusions can be drawn. First, the
simple luminosity evolution scenario for individual sources
outlined in Sect.~\ref{sect:scenario} appears to be fully
consistent with the data. We note however that, although the
formalism stems from the self-similar evolution models by Fanti et
al. (1995) and Begelman (1996, 1999), according to which both the
parameters $\eta$ and $\lambda$ are determined by the slope $n$ of
the density profile of the ambient medium, the data can be
satisfactorily fitted only if the two parameters are treated as
independent. This is not surprising, since self-similarity
can be easily broken under realistic conditions, and indeed
deviations from self-similarity were found in the two-dimensional
hydrodynamical simulations of Carvalho \& O'Dea (2002).

The fit is obtained for a positive value of $\eta$ and a negative
value of $p$, implying a decrease of the emitted power and of the
peak luminosity with source age or with decreasing peak frequency,
at variance with the Snellen et al. (2000) model. On the other
hand, our analysis confirms the rather flat slope of the
luminosity function, found by Snellen et al. (2000) who also
report indications of a high luminosity break, not required by the
data sets we have used. Snellen et al. (2000) argue that, in the
framework of a scenario whereby GPS sources increase their
luminosity until they reach a size $\sim 1\,$kpc and dim
thereafter, during the CSS and extended radio source phases, the
luminosity function of GPS galaxies can evolve into that extended
radio sources. Our results suggest that the GPS galaxies are the
precursors of extended radio sources with luminosities below the
break of the luminosity function ($L_{\rm break}\sim
10^{33}\,\hbox{erg}\, \hbox{s}^{-1}\,\hbox{Hz}^{-1}$ at 2.7 GHz,
cf. Dunlop \& Peacock 1990, De Zotti et al. 2005). Our best fit
model implies that the source luminosity is, roughly, inversely
proportional to the source age. If GPS sources have typical ages
of $\simeq 10^3\,$yr and extended radio sources of $\simeq
10^7\,$yr, we expect the latter to be $\sim 10^4$ times less
powerful than the former; as shown by Fig.~\ref{nup_Sp_Lp}, the
maximum value of the peak luminosity we found is $\sim
10^{36}\,\hbox{erg}\, \hbox{s}^{-1}\,\hbox{Hz}^{-1}$ (but this may
be a lower limit, since we have arbitrarily set the maximum
redshift at $z_f=1.5$), i.e. about 3 orders of magnitude larger
than $L_{\rm break}$. The GPS sources in the samples considered
here are thus expected to evolve into large-scale radio galaxies
with $L<L_{\rm break}$.

It must be stressed, however, that the uncertainties are very
large, so that firm conclusions must await for larger samples and
more complete redshift information. In particular, we have
checked that still acceptable fits can be obtained setting the
exponent $\eta$ of the relationship between the emitted radio
power and the source age to the value implied by Begelman's model
for $n=2$, i.e. $\eta=0.5$ (the total $\chi^2$ increases by
$\delta\chi^2\simeq 2$). In this case, the source luminosities
would typically decrease by a factor of $\sim 100$ from the GPS to
the extended radio source phase, and GPS sources could be the
progenitors of extended radio sources with luminosities both above
and below $L_{\rm break}$. It is plausible that sources have a
distribution of values of $\eta$ whereby the lower values become
increasingly rare, to account for the steepening of the luminosity
function above $L_{\rm break}$. Alternatively, the fraction of
prematurely fading sources may increase with luminosity. If
$\eta=0.5$, the best fit value of $\lambda$ is 1.02, and the peak
luminosity, $L_p(\nu_p)$ {\it increases}, albeit slowly, with {\it
decreasing} peak frequency [$L_p(\nu_p)\propto \nu_p^{-0.25}$, cf.
eq.~(\ref{eq:Lp})]. The other values of the parameters keep values
very close to those in Table~\ref{parameters}.

Another key observable quantity potentially providing crucial
constraints on evolutionary models is the linear size. The
self-similar evolution models predict its dependence on source
age, given the density profile of the ambient medium, and on the
other fundamental observable, the radio power. However, as noted
above, we could not fit the data with strictly self-similar
models, so that we no longer have a well defined relationship
between radio power and linear size. Such a relationship, however,
can be recovered through the {\it observed} relationship between
turnover frequency and linear size, $\nu_p \propto l^{-\delta}$,
with $\delta \simeq 0.65$; we have $L \propto
l^{-\eta\delta/\lambda}$. The anticorrelation between the latter
quantities implies that, since the radio power {\it decreases}
with decreasing peak frequency, it decreases with increasing
linear size. Once again, however, the uncertainties are very
large. The formal best fit values of the parameters would imply an
unrealistically steep decrease of the radio power, $L$, with
increasing linear size, $l$, but, if $\eta=0.5$, $L \propto
l^{-0.32}$ and $L_p(\nu_p)\propto \l^{p\delta/\lambda}\propto
l^{0.17}$.

In turn, fixing the exponent $\lambda$ of the relationship between
peak frequency and  source age to the value implied by Begelman's
model for $n=2$, i.e. $\lambda=\delta=0.65$, we get again an
increase of the minimum $\chi^2$ by $\delta\chi^2\simeq 2$, the
best fit value of $\eta$ is 0.76, while the other parameters are
essentially unchanged. In this case we have $L_p(\nu_p)\propto
\nu_p^{0.42}$, $L \propto l^{-0.76}$, and $L_p(\nu_p)\propto
l^{-0.27}$.

These examples illustrate the importance of the determination of
linear sizes for large complete samples of GPS sources. Given
their extreme compactness and their redshift distribution, images
with milli-arcsec resolution are generally required. It is thus
not surprising the the currently available information is scanty
and inhomogeneous. New VLBA data are however being acquired and
analyzed, for example on the HFP sample (Orienti et al. 2005). 
Future evolutionary models should comply also with the distributions of
linear sizes and with relationships of sizes with radio power.

Complementary information on the proposed evolutionary sequence
linking GPS sources to large scale radio sources is provided by
CSS sources, which could correspond to the intermediate phase when
sources have expanded out of the narrow line region but are still
within the host galaxy. More complete evolutionary studies should
take into account also these sources.

\begin{acknowledgements}
We warmly thank Daniele Dallacasa for useful discussions and
suggestions. This work has made use of the NASA/IPAC Extragalactic
Database NED which is operated by the JPL, California Institute of
Technology, under contract with the National Aeronautics and Space
Administration. The authors acknowledge financial support from the
Italian ASI and MIUR.
\end{acknowledgements}

\end{document}